\documentclass[11pt,fleqn,runningheads]{llncsToc}
\usepackage{latexsym,ifthen,amssymb,array,graphicx,psfrag,euscript,}
\usepackage[latin1]{inputenc}
\usepackage{a4wide}
\newcommand{\And}{\mathrel{\wedge}} 
\newcommand{\Or}{\mathrel{\vee}} 
\newcommand{\False}{\mti{false}} 
\newcommand{\True}{\mti{true}} 
\newcommand{\Imp}{\Rightarrow} 
\newcommand{\Follows}{\Leftarrow}
 
\newcommand{\Rel}{\leftrightarrow}

\newcommand{\Totfunc}{\rightarrow}

\newcommand{\Com}[2]{\mbox{\Esp{-1em}$#1$\Esp{1em}%
    \ifthenelse{\equal{#2}{ }}{}{\makebox[.9\textwidth]{\mbox{}\hfill$\{$ #2 $\}$}}}}
\newenvironment{Preuve}{\[\begin{array}{>{$}p{.9\textwidth}<{$}}}{\end{array}\]} 

\newcommand{\mti}[1]{\mbox{{\it #1}}}

\newcommand{\open}{\;\widehat{\,}\;}

\newcommand{\cpl}{\overline}
\newcommand{\ets}{\varnothing}

\newenvironment{PreuveL}{\[\begin{array}{>{\mbox{\Esp{-2ex}}\hfill}%
p{2.5ex}<{.\Esp{1ex}}>{\Esp{-1ex}$}p{.66\textwidth}<{$}>{;\ }p{.25\textwidth}}}{
\end{array}\]}

\newcommand{\Nat}{\mathbb{N}}
\newcommand{\Pow}{\mathbb{P}}

\newcommand{\card}{\mbox{\textsf{card}}}

\newcommand{\dom}{\mbox{\textsf{dom}}}

\newcommand{\fix}{\mbox{\textsf{fix}}}
\newcommand{\FIX}{\mbox{\textsf{FIX}}}
\newcommand{\pre}{\mbox{\textsf{pre}}}

\newcommand{\prd}{\mbox{\textsf{prd}}}

\newcommand{\grs}{\mbox{\textsf{grd}}}
\newcommand{\grd}{\mathit{grd}}

\newcommand{\Rarrow}{\Longrightarrow}

\newcommand{\bop}{}
\newcommand{\eop}{\mbox{}\hfill $\Box$}

\renewcommand{\SS}{\mathfrak S}
\newcommand{\Skip}{\mti{skip}}
\newcommand{\Sel}{\mathrel{[\mbox{\hspace{-.25ex}}]}} 
\newcommand{\sco}{\ensuremath{\mathrel{\mbox{\bf ;}}}}

\newcommand{\m}[1]{\ensuremath{\mathit{#1}}}
\newcommand{\Esp}[1]{\mbox{\hspace*{#1}}}
\newenvironment{PreuveCL}[1]{\[\begin{array}{>{\mbox{\Esp{-2ex}}\hfill}%
       p{2.5ex}<{.\Esp{1ex}}>{\Esp{-1ex}$}p{.66\textwidth}<{$}>%
       {\ifthenelse{\equal{#1}{N}}{ }{;\ }}%
       p{.25\textwidth}}}{\end{array}\]}
\newcommand{\dtl}{\mathrel{\triangledown}}
\newcommand{\B}{{\sf B}}
\newcommand{\Li}{{\cal L}}

\renewcommand{\bop}{}
\renewcommand{\eop}{\mbox{}\hfill $\square$}

\newcommand{\F}{{\cal F}}
\newcommand{\Fm}{\F_{m}}
\newcommand{\Fw}{\F_{w}}
\renewcommand{\L}{\EuScript{L}}
\newcommand{\Lm}{\L_{m}}
\newcommand{\Lw}{\L_{w}}
\newcommand{\Em}{{\cal E}_{m}}
\newcommand{\Ew}{{\cal E}_{w}}
\newcommand{\E}{{\cal E}}
\newcommand{\Ee}{\EuScript{E}} 
\newcommand{\T}{{\cal T}}
\newcommand{\Tw}{{\cal T}_{w}}
\newcommand{\Tm}{{\cal T}_{m}}
\renewcommand{\S}{{\cal S}}

\renewcommand{\a}{\alpha}
\renewcommand{\b}{\beta}
\newcommand{\Pred}{\mbox{\sf Pred}}
\renewcommand{\P}{{\cal P}}
\newcommand{\Q}{{\cal Q}}

\newcommand{\sub}{\mbox{\sf sub}}
\titlerunning{A Fixpoint Semantics of Event Systems}
\begin{document}
\thispagestyle{empty}
\parbox{.8\textwidth}{ %
  {\Huge\bf IMAG}\\\\
  {\large\bf Institut d'Informatique et de \\[.5ex]
    Math\'{e}matiques Appliqu\'{e}es \\[.5ex]
    de Grenoble} 
}\hfill
\vspace{3cm}
\parbox{\textwidth}{%
  \begin{center}
    {\Huge\bf LSR}\\[1ex]
    {\LARGE\bf Laboratoire Logiciels, Syst\`{e}mes, R\'{e}seaux}
  \end{center}
}
\vfill
\begin{minipage}{\textwidth}
\title{%
  {\LARGE\bf RAPPORT DE RECHERCHE}\\[3ex]
  A Fixpoint Semantics of Event Systems with and without Fairness
  Assumptions
  }
\author{H\'{e}ctor Ru\'{\i}z Barradas\inst{1,2} \and Didier Bert\inst{2}}
 \institute{Universidad Aut\'{o}noma Metropolitana Azcapotzalco, M\'{e}xico
   D. F.,  M\'{e}xico\\
 \email{hrb@correo.azc.uam.mx, Hector.Ruiz@imag.fr}
 \and
 Laboratoire Logiciels, Syst\`{e}mes, R\'{e}seaux - 
 LSR-IMAG - Grenoble, France\\
 \email{Didier.Bert@imag.fr}
 }
\maketitle
\end{minipage}
\vfill
\parbox{\textwidth}{%
RR  1081-I LSR 21 \hfill D\'{e}cembre 2005 \\[3ex]
\begin{center}
  {B.P. 72 - 38402 SAINT MARTIN D'HERES CEDEX - France}\\[2ex]
  {Centre National de la Recherche Scientifique}\\[1ex]
  {Institut National Polytechnique de Grenoble}\\[1ex]
  {Universit\'e Joseph Fourier Grenoble I}
\end{center}}
\newpage
\mbox{}
\newpage
\thispagestyle{empty}
\newpage
\thispagestyle{empty}
\begin{center}
\Large\bf A Fixpoint Semantics of Event Systems with and without Fairness
  Assumptions
\end{center}
\vfill
\section*{Abstract}
  We present a fixpoint semantics of event systems. The semantics is
  presented in a general framework without concerns of
  fairness. Soundness and completeness of rules for deriving
  \emph{leads-to\/} properties are proved in this general
  framework. The general framework is instantiated to minimal progress
  and weak fairness assumptions and similar results are obtained. We
  show the power of these results by deriving sufficient
  conditions for \emph{leads-to} under minimal progress proving
  soundness of proof obligations without reasoning over state-traces.

\subsection*{Keywords}
Liveness properties, event systems, action systems, 
{\sc unity} logic, fairness, weak fairness, minimal progress, set
transformer, fixpoints.

\section*{R\'esum\'e}
Dans ce rapport nous pr\'esentons une s\'emantique de point fixe pour
les syst\`emes d'\'ev\'enements. La s\'emantique est pr\'esent\'ee
dans un cadre g\'en\'erale sans consid\'erations d'\'equit\'e. La 
coh\'erence et la compl\'etude des r\`egles pour d\'eriver des
propri\'et\'es \emph{leads-to\/} est prouv\'ee dans ce
cadre g\'en\'eral. Le cadre g\'en\'eral est instanci\'e avec des
hypoth\`eses de progr\`es minimal et d'\'equit\'e faible, et des 
r\'esultats similaires sont prouv\'es. Nous montrons la puissance de 
ces r\'esultats par la d\'erivation de conditions suffisantes pour
des propri\'et\'es \emph{leads-to} sous l'hypoth\`ese de progr\`es
minimal, et nous prouvons la coh\'erence de ces r\`egles sans
raisonner sur les traces d'\'etats.

\subsection*{Mots-cl\'es}
Propri\'et\'es de vivacit\'e, syst\`eme d'\'ev\'enements, syst\`emes
d'actions, logique {\sc unity}, \'equit\'e, \'equit\'e faible, progr\`es 
minimal, transformateurs d'ensembles, point fixes.
\vfill
\mbox{}
\newpage
\mbox{}
\newpage
\setcounter{tocdepth}{2}
\mbox{}
\vfill
\begin{minipage}{\textwidth}
\tableofcontents
\end{minipage}
\vfill
\mbox{}
\newpage
\mbox{}
\newpage
\pagestyle{plain}
\setcounter{page}{1}


\section{Introduction}\label{intro}
Action systems, or event systems, are useful abstractions to model
discrete systems. Many formalisms have been proposed to model action
systems. In these formalisms, the behavior of a system is described in
terms of observations about its state, and they are known as
\emph{state based\/} formalisms. As examples of state based formalisms
we can cite Back's action system formalism \cite{BaKuDPCC} and {\sc
  unity} \cite{ChMiPPD}. All of these formalisms have a common aspect:
their semantics is founded on state-traces of transition systems.

State traces of transitions systems impose an operational reasoning
about the behavior of a system. However this operational behavior can
be hidden by using temporal logic to specify safety and liveness
properties. Semantics of temporal formulas is given by state-traces of
transition systems. A proof system allows us to derive properties
from other proved properties without an operational reasoning, only
by symbolic calculations. Soundness and completeness of the logic are
established by proofs relating logical formulas with assertions about
state-traces. So at this point, we come back to operational reasoning
about the transition systems.

A more abstract possibility to define the semantics of action systems
is to base it on fixpoints of set (or predicates) transformers.
Inspired from \cite{AbrMusIDCB} and \cite{RaJuMDPR}, we characterize
certain liveness properties as fixpoints of set transformers modeling
iteration of events, under minimal progress or weak fairness
assumptions. We are only interested in properties of type $P$
\emph{leads-to} $Q$, where $P$ and $Q$ are predicates on the state
space of a system, with the informal meaning: ``the system \emph{reaches\/}
a state satisfying $Q$ when its execution arrives at any state in
$P$''. The fixpoint characterizing this property denotes the largest
subset of states, containing all states satisfying $P$, where
\emph{termination\/} of iteration of events in the system, is
guaranteed to terminate in a state satisfying $Q$. Soundness and
completeness of rules allowing derivation of \emph{leads-to\/}
properties is proved by demonstrating that notions of reachability and
termination are equals under minimal progress or weak fairness.
Moreover, we give two examples of applications of these results: The
first one is a proof of sufficient conditions for liveness properties under
minimal progress given in \cite{AbrMusIDCB}. The second one is an
original result which gives sufficient conditions to derive a liveness
property under minimal progress when the given property holds under
weak fairness.

This report is an extended version of the semantics of event systems
presented in \cite{RuBeFSES}. In particular all proofs of that paper
are given here in an explicit way, and a new section, considering the
semantics  with the strongest invariant, is presented. Paper
\cite{RuBeFSES} presents comparisons with other works dealing with
fairness properties. This part is not given here. The report
is structured as follows. In Section \ref{str-unity}, we
present a system as a set transformer and we give syntax and semantics
of common set transformers used to model events, or actions, in the
system. Moreover we give a brief review of liveness properties in {\sc
  unity} logic to specify properties of an event system. In Section
\ref{reachterm}, we develop our semantics of event systems and we prove
equality (soundness and completeness) between notions of termination
and reachability. In Section \ref{deriving}, we give examples of
sufficient conditions to derive liveness properties using the results
of the previous section. Finally we give our conclusions and future
work in Section \ref{conc}. Annex A presents the proof of the
\emph{leads-to\/} properties as a relation between predicates or
sets. Annex B shows the extension of the semantics to consider the
strongest invariant. Annexes C, D and E present the proofs of sections
3 and 4. 

\section{Set Transformers and {\sc unity} Logic in Event Systems}
\label{str-unity}

In this section we introduce the main considerations about event
systems and the specification of liveness properties in {\sc unity}
logic. This section is divided in two parts. The first part presents an
event system as a set transformer and introduces the notion of
liberal set transformer, as well as the dovetail operator that is used
to model a weak fairness assumption. In the second part we recall the
main ideas in the specification and proof of liveness properties under
two fairness assumptions in {\sc unity}-like logic.

\subsection{Set Transformers}\label{set-trans}
A set transformer is a total function of type $\Pow({\cal U})\Totfunc
\Pow({\cal U})$ for a certain set ${\cal U}$. An event system is made
out of a family of events. Any event may be executed in any state
where its \emph{guard\/}, boolean condition on the state, holds. When
the guard of an event holds, we say that the event is enabled. As in
event-\B\ systems, we considered a system with state variable $x$ and
invariant $I$. The state space $u$ of the system is the set of states
where $I$ holds: $u=\{\,z\,|\,I(z)\,\}$. Therefore events in the
system are modeled by conjunctive set transformers $E_i$ of type
$\Pow(u) \Totfunc\Pow(u)$, where $i$ belongs to certain finite index
set $L$.  Consequently, the system is modeled by a conjunctive set
transformer $S$ which is the bounded choice of events $E_i$: $S =
\Sel_{i\in L} E_{i}$. We denote by $\S$ the set of events in $S$:
$\S=\{E_{i}\;|\;i\in L\}$.

For any set transformer $T$ of type $\Pow(u)\Totfunc\Pow(u)$ and
subset $r$ of $u$, $T(r)$ denotes the largest subset of states where
execution of $T$ must begin in order for $T$ to terminate in a state
belonging to $r$ \cite{AbrTBB}.  Primitive set transformers considered
in this paper are similar to the primitive generalized substitutions in
\B: skip, bounded choice, sequence, guarded and conditioned set
transformer. Following the work reported in \cite{RuBePOWF}, for any
set transformer $T$, and subset $r$ of $u$, we denote by $\Li(T)(r)$
the \emph{liberal set transformer} of $T$, which denotes the largest
subset of states where the execution of $S$ must begin in order for
$T$ to terminate in a state belonging to $r$ or loop. Common set
transformers and liberal set transformers are defined as
follows:\footnote{%
  For any set transformer $T$, $(\Li)(T)(r)$ denotes definition for
  the set $T(r)$ or the set $\Li(T)(r)$.}
\[
\hspace{-1\mathindent}
\parbox{\textwidth}{
  \begin{tabular}{ll}
  \parbox{.5\textwidth}{
    \[\begin{array}{l}
      (\Li)(\Skip)(r) = r \\
      (\Li)(F\Sel G)(r) =(\Li)(F)(r)\cap (\Li)(G)(r)\\
    \end{array}\]
  }
  &
    \parbox{.5\textwidth}{
    \[\begin{array}{l}
      (\Li)(F\sco G)(r) = (\Li)(F)((\Li)(G)(r)) \\
      (\Li)(p \Rarrow F)(r) = \cpl{p}\cup (\Li)(F)(r) 
    \end{array}\]
    }
\end{tabular}}
\]
In the guarded event, $\cpl{p}$ denotes $u-p$. For the preconditioned
event we have:
 \[\begin{array}{l}
      (p\;|\;F)(r) = p\cap F(r) \\
      \Li(p\;|\;F)(r) = \left\{
        \begin{array}{ll}
         p\cap\Li(F)(r) &  \mbox{if $r\not = u$} \\
         \Li(F)(r) &  \mbox{if $r= u$}
        \end{array}
      \right.
    \end{array}\]
Definitions of liberal set
transformers presented here are the set counterpart of definitions in
\cite{SduIBRB}. The set transformers $F(r)$ and $\Li(F)(r)$ for event $F$ and
postcondition $r$ are related by the pairing condition: 
\[F(r) = \Li(F)(r) \cap \pre(F)\]
where $\pre(F)$, the termination set of $F$, is equal to $F(u)$.  From
the pairing condition, we conclude: \[F(u)=u \Imp F(r) = \Li(F)(r)\]
We say that a set transformer $F$ is \emph{strict\/} when it respects the
excluded miracle law: \[F(\ets)= \ets\]

\noindent For any set transformer $F$, when $F(r)$ or $\Li(F)(r)$ are
recursively defined: 
\[F(r) = {\cal F}(F(r)) \quad \mbox{or} \quad \Li(F)(r) = {\cal
  G}(\Li(F)(r))\]
for monotonic functions ${\cal F}$ and ${\cal G}$,
according to \cite{HehDCO} we take $F(r)$ as the strongest solution of
the equation $X = {\cal F}(X)$ and $\Li(F)(r)$ as the weakest solution
of the equation $X = {\cal G}(X)$. As these solutions are fixpoints,
we take $F(r)$ as the least fixpoint of ${\cal F}$ ($\fix({\cal F})$)
and $\Li(F)(r)$ as the greatest fixpoint of ${\cal G}$ ($\FIX({\cal
  G})$).

\subsubsection*{The Dovetail Operator}
To model a weak fairness assumption, we use the dovetail operator
$\dtl$ \cite{BrNeAFCD}, which is a fair nondeterministic choice
operator. The dovetail operator is used to model the notion of fair
scheduling of two activities. Let $A$ and $B$ be these activities,
then the operational meaning of the construct $ A \dtl B$ denotes the
execution of commands $A$ and $B$ fairly in parallel, on separate
copies of the state, accepting as an outcome any proper, nonlooping,
outcome of either $A$ or $B$.  The fair execution of $A$ and $B$ means
that neither computation is permanently neglected if favor of the
other. 

The semantic definition for dovetail operator in \cite{BrNeAFCD} is given
by definition of its weakest liberal precondition predicate transformer
($\m{wlp}$) and its termination predicate $\m{hlt}$. We give an equivalent
definition using the weakest liberal set transformer $\Li$ and its
termination set $\pre$:
\begin{eqnarray}
  \label{Ldtl}
  \lefteqn{\Li(F\dtl G)(r) = \Li(F)(r)\cap\Li(G)(r)}\\
  \label{predtl}
  \lefteqn{\pre(F\dtl G)  =  (F(u)\cup G(u))\cap (\cpl{F(\ets)} \cup G(u))
    \cap (\cpl{G(\ets)}\cup F(u))}
\end{eqnarray}
We remember that $\grs(F) = \cpl{F(\ets)}$. From these definitions, in
\cite{RuBePOWF} we prove the guard property of the dovetail: $\grs(A\dtl B) =
\grs(A) \cup \grs(B)$. 

A motivating example of the use of the dovetail operator is given in
\cite{BrNeAFCD}. In that example the recursive definition:
$ X = (n:=0 \dtl (X\sco n:=n+1))$ 
which has as solution ``set $n$ to any natural number'', is contrasted
with the recursion $Y = (n:=0 \Sel (Y\sco n:=n+1))$ 
 which has as solution ``set $n$ to any natural number or loop''. The
possibility of loop in $X$ is excluded with the dovetail operator because
the fair choice of   statement $n:=0$ will certainly occur. In $Y$ the
execution of that statement is not ensured.

\subsection{Liveness Properties in event systems}\label{live}
In this section we give a brief summary of some results in the
specification and proof of liveness properties presented in \cite{RuBeSPLP},
\cite{RuBePDHE} and \cite{RuBePOWF}. In these works, we propose the
use of {\sc unity} logic to specify and prove liveness properties in
event-\B\ systems. 

Liveness properties are divided in two groups: basic and general
liveness properties. Each one of these properties are specified by
relations on the state of the system. In order to specify and prove
these properties we consider a minimal progress or a weak fairness
assumption.

\subsubsection*{Basic Properties under Weak Fairness}
A weak fairness assumption states that any continuously enabled event is
infinitely often executed. For any event $G$ in the set $\S$, we write $G \cdot
P \gg_{w} Q$ (pronounce ``by event $G$, $P$ ensures $Q$'') to specify
that by the execution of event $G$ in a state satisfying $P$ the
system goes to another state satisfying $Q$, under a weak fairness
assumption. In \cite{RuBePDHE} we propose sufficient conditions WF0
and WF1, to guarantee the intended meaning of these properties. These
conditions were stated in terms of predicates, but we present them as
set expression:
\begin{equation}
  \label{wf}
  p\cap \cpl{q}\subseteq S(p\cup q)\cap \grs(G)\cap G(q)\Imp G\cdot
  x\in p\gg_{w} x\in q 
\end{equation}
where $x$ is the state variable of $S$, $p=\{z|z\in u \And P\}$ and
$q=\{z|z\in u \And Q\}$ for certain predicates $P$ and $Q$.
 
\subsubsection*{Basic Properties under Minimal Progress}
In a minimal progress assumption, if two or more statements are
enabled in a given state, the selection of the statement enabled for
execution is non-deterministic. We write $P \gg_{m} Q$ (pronounce
``$P$ ensures $Q$'') to specify that execution of any event of $S$, in
a state satisfying $P$, terminates into a state establishing $Q$.  In
\cite{RuBeSPLP} we give sufficient conditions MP0 and MP1 to prove basic
properties under minimal progress. We present them as a set expression
as follows, for sets $p$ and $q$ defined as above:
\begin{equation}
  \label{mp}
  p\cap \cpl{q}\subseteq S(q)\cap \grs(S)\Imp x\in p\gg_{m} x\in q
\end{equation}
\subsubsection*{General Properties}\label{general}
General liveness properties are specified by the \emph{leads-to}
operator $\leadsto$. Depending on the fairness assumption considered,
we have general liveness properties under minimal progress or weak
fairness assumptions. However, the \emph{leads-to\/} relation is
defined in the same way as the closure relation, containing the base
relation and it is both transitive and disjunctive.  A property
$P\leadsto Q$ holds in an event system, if it is derived by a finite
number of applications of the rules defined by the {\sc unity} theory:
\begin{center}
\begin{tabular}{|c|l|l|}\hline
 &\ ANTECEDENT &\ ~~CONSEQUENT \\ \hline
& & \\[-2mm]
\textbf{~~BRL~~~}&\ $P \gg Q$ &\ $P \leadsto Q$ \\[1mm]
\textbf{~TRA~} &\ $P\leadsto R$, $R\leadsto Q$ &\ $P \leadsto Q$ \\[1mm]
\textbf{~DSJ~} &\ $\forall i \cdot (i\in I \Imp P(i)\leadsto Q)~~$&\  $\exists
i \cdot (i\in I \And P(i))\leadsto Q$~~~~ \\[2mm]
\hline
\end{tabular}
\end{center}
$P \gg Q$, in the BRL rule stands for the basic liveness property
$G\cdot P \gg_{w} Q$ for some $G$ in $\S$ in case where we consider a
property under a weak fairness assumption or $P \gg_{m} Q$, in the case
where we consider a minimal progress assumption. In the
disjunction rule DSJ, $I$ is any index set.

\section{Reachability and Termination}\label{reachterm}

In this section, we prove soundness and (relative) completeness of
rules BRL, TRA and DSJ for general liveness properties under minimal
progress and weak fairness assumptions in event systems. These rules
are sound if for any property $P\leadsto Q$, iteration of events,
under minimal progress or weak fairness assumptions, starting in a
state satisfying $P$, leads to a state in the system where $Q$ holds.
Completeness of these rules is proved by showing that $P\leadsto Q$
can be derived from the fact that any iteration of events, starting in
a state where $P$ holds, terminates into a state satisfying $Q$.

We do not expect that any iteration of events in a system terminates
into a state where the guards of every event are disabled. However we
can model an iteration of events which \emph{always\/} terminates in a
certain state by supposing, just for the reasoning, that the events in
the system are embedded in a certain guarded event which models the
iteration under a fairness assumption. The iteration only proceeds
when the guard of that event is enabled. \emph{Termination\/} of the
iteration will be in a state where the guard does not hold. In this
way, if the guard of the iteration is $\neg Q$, and the iteration
starts in a state where $P$ holds, the system reaches a state where
$Q$ holds. \emph{Reachability\/} from $P$ to $Q$ is then associated to
termination of the iteration of events. In the following subsection,
we formalize our claims in a general framework without concerns of
fairness, and then we particularize these results to minimal progress
 or weak fairness assumptions in other two subsections.

To simplify matters, the strongest invariant \cite{SanESAU} is not
considered in definitions of this section. Therefore, instead of
implications in proof obligations (\ref{wf}) and (\ref{mp}) used to
prove basic liveness properties under weak fairness or minimal
progress assumption respectively, we consider them as definitions. In
annex \ref{SI} we restate the results given in this section to
consider the strongest invariant and we consider again, proof
obligations (\ref{wf}) and (\ref{mp}) as implications, as they are
stated. 

\subsection{A General Framework}\label{Gsouncomp}
In this subsection we define a set transformer to model iteration of
events and we state its main characteristics. We use this set
transformer to define the \emph{termination\/} relation. Then we give
a representation of \emph{leads to\/} relation in {\sc unity} logic as
a relation between subsets of $u$ and we use it to define the
\emph{reachability\/} relation. Finally we prove that the
\emph{termination\/} and the \emph{reachability\/} relations are
equal.

\subsubsection{Termination}\label{ter}
We consider a set transformer $W$ which models a \emph{step\/} of the
iteration of events in a system $S$. At this time we cannot define the
meaning of such a step, however we need two properties of $W$: it must
be \emph{monotonic\/} and \emph{strict}. When we particularize the
iteration under a fairness assumption, the meaning of $W$ will be
given in terms of $S$. For any $r$ in $\Pow(u)$, $W(r)$ denotes the
largest subset of states where the execution of $W$ must begin in
order for $W$ to terminate in a state belonging to $r$.  

To model the iteration of events until the system reaches a state in a
certain set $r$ in $\Pow(u)$, we define a guarded event $\F(r)$:
\begin{equation}
  \label{defF}
  \F(r) = (\cpl{r}\Rarrow W)
\end{equation}
for any $r\in\Pow(u)$, which allows
iteration of $W$ when the system stays in any state in $\cpl{r}$.
Iteration of $\F(r)$ is modeled by the $\open$ operator $\F(r)\open$.
As this operator has a recursive definition:
\[ \F(r)\open = (\F(r) \sco \F(r)\open) \Sel \Skip\] 
the set where termination of $\F(r)\open$ is
guaranteed ($\pre(\F(r)\open)$) is given by $\fix(\F(r))$
\cite{AbrTBB}. 

As $W$ may model an unbounded non determinist set transformer, we use
the \emph{Generalized Limit Theorem\/} to formally justify that any
iteration of $\F(r)$ starting in $\pre(\F(r)\open)$ terminates in some
state of $r$. This theorem characterizes the least fixpoint of monotonic
functions as an infinite join. We use the version presented in
\cite{NelAGDC}, particularizing the theorem to monotonic set
transformers.  The theorem is as follows:
\begin{theorem}\label{tglim}
\emph{(Generalized Limit Theorem)}\\
Let $f$ be a monotonic set transformer, and let $f^\alpha$, for
ordinal $\alpha$, be defined inductively by
\begin{equation}\label{iterate}
 f^{\alpha} = \bigcup \beta\cdot (\beta < \alpha\;|\;f(f^{\beta}))
\end{equation}
Then $\fix(f) = f^{\alpha}$ for some ordinal $\alpha$. 
\end{theorem}
The proof of this theorem is given in \cite{NelAGDC}. It states that
we can choose any ordinal $\gamma$, such that $\gamma >
\card(\dom(f))$, and then we must have $f^{\alpha}=f^{\beta}$ for some
$\alpha < \beta < \gamma$. Then it is proved that the common value of
$f^{\alpha}$ and $f^{\beta}$ is the least fixpoint of $f$.

As $W$ is a monotonic function, $\F(r)$ (\ref{defF}) is monotonic, and
theorem \ref{tglim} can be applied  to calculate the least fixpoint of
$\F(r)$. According to the theorem, we conclude that $\F(r)^{0} = \ets$
and $\F(r)^{1}= r$ \emph{because\/} $W$ is strict. Moreover, for any
ordinal $\a$, $\F(r)^{\a+1} =\F(r)(\F(r)^{\a})$ and $\F(r)^{\a}
\subseteq \F(r)^{\a+1}$. This fact formally supports our claim that the
termination set of $\F(r)\open$, contains states where any iteration
of $\F(r)$ terminates in a state into $r$. Now, we can define the
\emph{termination\/} relation $\T$ as follows:
\begin{definition}\label{defT}
\emph{(Termination Relation)}
\begin{eqnarray}
\label{term}
\lefteqn{\T=\{\,a\mapsto b\,|\,a\subseteq u\And b\subseteq u\And
  a\subseteq \fix(\F(b))\,\}} 
\end{eqnarray}
\end{definition}

\subsubsection{Reachability}\label{reachsec}
As presented in section \ref{general}, \emph{leads-to\/} relation of
{\sc unity} logic is defined as a relation between predicates on the
state of programs. In this section we define a similar relation, $\L$,
but instead of predicates, we define it as a relation between subsets
of states in $u$ ($\L\subseteq \Pow(u)\times\Pow(u)$). Any pair $a\mapsto b$ in
$\L$ indicates that the system reaches a state in $b$, when its
execution arrives at any state in $a$. For this reason we name $\L$ as
the \emph{reachability\/} relation.

Definition of $\L$ is given by induction. The base case needs
definition of the basic relation $\E$. At this time
$\E$ cannot be defined. As indicated in section \ref{live}, basic
liveness properties depend on fairness assumptions. $\E$ will be
defined in the following sections according to minimal progress or
weak fairness assumptions. However, these definitions must satisfy two
requirements.  The first requirement is as follows: If $a\subseteq b$,
for any $a$ and $b$ in $\Pow(u)$, then $a\mapsto b \in \E$ must hold.
The second requirement relates $\E$ with the set transformer $W$: For
any ordered pair $a\mapsto b\in\E$, the inclusion $a \cap \cpl{b}
\subseteq W(b)$ must hold. This inclusion indicates that any execution
of $W$ starting in $a\cap \cpl{b}$, terminates into a state of $b$.

\begin{definition}\label{reach}
  \emph{(Reachability Relation)}\\
  The \emph{reachability} relation $\L$, $\L\in \Pow(u)\Rel\Pow(u)$,
  is defined by the following
  induction scheme:\\[.5ex]
\noindent\emph{(\textbf{SBR})}: $\E\subseteq \L$\\[.25ex]
\emph{(\textbf{STR})}: $\L \!\sco \L \subseteq \L$\\[.25ex]
\emph{(\textbf{SDR})}: $\forall (q,l)\cdot (q\in\Pow(u)\And l\subseteq
\Pow(u) \Imp (l\times\{q\} \subseteq \L \Imp \bigcup(l)\mapsto
q \in \L))$ \\[.25ex]
\emph{\textbf{Closure}}: $\forall \m{l'}\cdot (l'\in u\Rel u \And
\E\subseteq \m{l'}\And \m{l'}\sco \m{l'}\subseteq \m{l'}\And$\\
$\forall (q,l)\cdot (q\in \Pow(u)\And l\subseteq \Pow(u)\And l\times
\{q\}\subseteq \m{l'}\Imp \bigcup(l)\mapsto q\in \m{l'})\Imp
\L\subseteq \m{l'})$
\end{definition}
$\bigcup(l)$ in the SDR rule and the closure clause, denotes the
generalized union of subsets in $l$. Rules SBR, STR and SDR are the
set counterpart of the basic rule for \emph{leads-to\/} BRL,
transitivity rule TRA and disjunction rule DSJ respectively, as
defined in section \ref{live}.

In order to connect $\L$ with the \emph{leads-to\/} relation of {\sc
    unity} logic, we have the following equivalence:
\begin{eqnarray}
\label{defequlto}
\lefteqn{P(x)\leadsto Q(x)\equiv \{\,z\,|\,z\in u\And P(z)\,\}\mapsto
  \{\,z\,|\,z\in u\And Q(z)\,\}\in \L} 
\end{eqnarray} 

We note that property $P\leadsto Q$ in {\sc unity} is equivalent to
$P\!\And\! I\leadsto Q\!\And\! I$, considering $I$ as an invariant
of~$S$, because the \emph{leads-to} relation is defined in states
reachable from the initial conditions \cite{SanESAU}. The proof of
this equivalence is given in annex \ref{equsetpred}.

\subsubsection{Soundness and Completeness}\label{souncomp}
We are now ready to state our main theorem, formally indicating that
\emph{termination\/} and \emph{reachability\/} relations are equal:
\begin{theorem}\label{Tsoundcomp}
  \emph{(Soundness and Completeness)}\\
  Let $W$ be a monotonic and strict set transformer and $\F(r) =
  (\cpl{r}\Rarrow W)$ for any $r$ in $\Pow(u)$. Let relations $\T$ and
  $\L$ be defined as definitions \ref{defT} and \ref{reach}
  respectively. Considering (a) $a\mapsto b\in \E\Imp a\cap
  \cpl{b}\subseteq W(b)$, (b) $a\subseteq b \Imp a\mapsto b\in\E$ and
  (c) $W(r)\mapsto r\in\L$, for any $a$, $b$ and $r$ in $\Pow(u)$, the
  following equality holds:
  \[\L = \T\]
\end{theorem}
Premise (a) and (b) were commented in the previous section. Premise
(c) asserts that any set $r$ is reached from the set $W(r)$ which is
the largest subset of states where a step of the iteration terminates
in $r$.

The proof of this theorem is given in two parts: first we prove the
inclusion $\L\subseteq \T$ and then $\T\subseteq \L$.

\paragraph{Proof of $\L\subseteq\T$} 
The proof of this inclusion follows from the closure clause in 
definition \ref{reach}, particularizing the quantified variable $l'$ to
relation $\T$. Then  $\L\subseteq\T$ follows from $\E\subseteq\T$, $\T
; \T \subseteq  \T$ and $l\times \{q\}\subseteq \T\Imp
\bigcup(l)\mapsto q\in \T$ for any $l$ in $\Pow(\Pow(u))$ and $q$ in
$\Pow(u)$. 

The proof of $\E\subseteq\T$ uses the following property for monotonic
function $f$ and iteration defined in (\ref{iterate}):
\begin{eqnarray}
\label{Pfix1}
\lefteqn{\forall \a\cdot (f^{\a}\subseteq \fix(f))}
\end{eqnarray} 
which is easily proved by transfinite induction; the proof is given in
appendix \ref{AppendSounComp}. The proof of $\E \subseteq \T$ is given by the
proof of $a\mapsto b\in \E\Imp a\mapsto b\in \T$:
\newcommand{\Fb}{\F(b)}
\begin{PreuveL}
1 & a\mapsto b\in \E &  premise\\
2 & a\cap \cpl{b}\subseteq W(b) &  1 and hyp. (a)\\
3 & a\subseteq \Fb(b) &  2 and def. (\ref{defF})\\
4 & a\subseteq \Fb^{2} &  3 and iterate (\ref{iterate})\\
5 & a\subseteq \fix(\Fb) &  4 and (\ref{Pfix1})\\
6 & a\mapsto b\in \T &  5 and def. (\ref{term})
\end{PreuveL}

In order to prove the transitivity of $\T$, we need the following
property: 
\begin{equation}\label{incfix}
a\mapsto b\in \T\Imp \fix(\F(a))\subseteq \fix(\F(b))
\end{equation}
for any $a$ and $b$ in $\Pow(u)$. Taking $a\mapsto b\in\T$ as a
premise, and considering $\fix(\F(a))$ as the least fixpoint of
$\F(a)$, in order to prove  property (\ref{incfix}) it suffices to prove
$\F(a)(\fix(\F(b)))\subseteq \fix(\F(b))$, which follows directly from
$a\mapsto b\in\T$ and $\F(b)(\fix(\F(b)))=\fix(\F(b))$. Now the proof
of $\T \sco \T\subseteq\T$ is equivalent to prove $a\mapsto b\in
\T\sco \T\Imp a\mapsto b\in \T$ for any $a$ and $b$ in $\Pow(u)$:
\begin{PreuveL}
1 & \exists c\cdot (a\mapsto c\in \T\And c\mapsto b\in \T) &  from
$a\mapsto b\in\T ;\T$ \\ 
2 & \exists c\cdot (a\subseteq \fix(\F(c))\And c\mapsto b\in \T) &  1
and def. $\T$\\ 
3 & \exists c\cdot (a\subseteq \fix(\F(c))\And \fix(\F(c))\subseteq
\fix(\F(b))) &  2 and (\ref{incfix}) \\ 
4 & a\subseteq \fix(\F(b)) &  3\\
5 & a\mapsto b\in \T &  6 and def. $\T$
\end{PreuveL}
Finally, the proof of $l\times \{q\}\subseteq \T\Imp \bigcup(l)\mapsto
q\in \T$ is as follows:
\begin{PreuveL}
1 & l\times \{q\}\subseteq \T &  premise\\
2 & \forall p\cdot (p\in l\Imp p\mapsto q\in \T) &  1\\
3 & \forall p\cdot (p\in l\Imp p\subseteq \fix(\F(q))) &  2 and def. $\T$\\
4 & \bigcup(l)\subseteq \fix(\F(q)) &  3\\
5 & \bigcup(l)\mapsto q\in \T &  4 and def. $\T$
\end{PreuveL}
This last deduction concludes the proof of $\L\subseteq \T$.

\paragraph{Proof of $\T\subseteq\L$}
The proof of this inclusion requires the following property:
\begin{eqnarray}
\label{ordprop}
\lefteqn{\forall r\cdot (r\in \Pow(u)\Imp \F(r)^{\alpha}\mapsto r\in
  \L) \quad \mbox{for any ordinal $\a$}}
\end{eqnarray}
The proof of (\ref{ordprop}) is done by transfinite induction.  For a
successor ordinal we need to prove $\F(r)^{\a}\mapsto r\in \L \Imp
\F(r)^{\a+1}\mapsto r\in \L$; this proof is given in appendix
\ref{AppendSounComp}. For a limit ordinal we prove
$\forall \b\cdot (\b< \a\Imp \F(r)^{\b}\mapsto r\in \L) \Imp
\F(r)^{\a}\mapsto r\in \L$:
\begin{PreuveL}
1 & \forall \b\cdot (\b< \a\Imp \F(r)^{\b}\mapsto r\in \L) &  ind. hyp.\\
2 & \forall \b\cdot (\b< \a\Imp W(\F(r)^{\b})\mapsto \F(r)^{\b}\in \L)
&  from hyp. (c)\\ 
3 & \forall \b\cdot (\b< \a\Imp W(\F(r)^{\b})\mapsto r\in \L) &  2, 1 and STR\\
4 & r\mapsto r\in \L &  hyp. (b) and SBR\\
5 & \forall \b\cdot (\b< \a\Imp r\cup W(\F(r)^{\b})\mapsto r\in \L) &
4, 3 and SDR\\ 
6 & \forall \b\cdot (\b< \a\Imp \F(r)(\F(r)^{\b})\mapsto r\in \L) &
def. $\F(r)$ and 5\\ 
7 & \bigcup \b\cdot (\b< \a\;|\;\F(r)(\F(r)^{\b}))\mapsto r\in \L &  6 and SDR\\
8 & \F(r)^{\a}\mapsto r\in \L &  7 and def. iterate
\end{PreuveL}

\noindent Using (\ref{ordprop}), we prove $\T\subseteq\L$ by the
proof of $a\mapsto b\in\T \Imp a\mapsto b\in\L$ for any $a$ and $b$
in $\Pow(u)$ as follows:
\begin{PreuveL}
1 & a\subseteq \fix(\F(b)) &  from $a\mapsto b\in\T$\\
2 & \exists \a\cdot (a\subseteq \F(b)^{\a}) &  1 and theorem \ref{tglim}\\
3 & \exists \a\cdot (a\mapsto \F(b)^{\a}\in \L) &  2, (b) and SBR\\
4 & \exists \a\cdot (a\mapsto \F(b)^{\a}\in \L\And \F(b)^{\a}\mapsto b\in
\L) &  3 and (\ref{ordprop})\\ 
5 & a\mapsto b\in \L &  4 and STR
\end{PreuveL}
This deduction concludes the proof of theorem \ref{Tsoundcomp}.

\subsection{Minimal Progress}\label{minimal}
In this paragraph we define the \emph{termination\/} and
\emph{reachability\/} relations under minimal progress and we prove
that they satisfy the premises of theorem \ref{Tsoundcomp}. Therefore
we claim that relations $\T$ and $\L$ are equal  in the case
of minimal progress. 

\subsubsection{Termination under MP}\label{termMP}
To model a step of the iteration of events of system $S$ under
minimal progress assumptions, we note that if we need to establish a
certain postcondition when this step is achieved, any event in $S$
must be able to establish the postcondition. Moreover, as we are
interested in the execution of any event, we need to start the
execution step in a state satisfying the guard of at least one
event. Therefore, taking into account these considerations, we propose
the following preconditioned set transformer:
\begin{equation}\label{Wmp}
 W_{m} = \grs(S)\;|\;S
\end{equation}

From definition of preconditioned set transformer in Section
\ref{set-trans} we actually have that $W_{m}(r) = \grs(S)\cap S(r)$. From
monotonicity of $S$ , we derive the monotonicity of $W_m$ and $W_{m}(\ets)
= (\grs(S)\cap S(\ets)) = \ets$ which proves the strictness of $W_m$.

The body of the iteration of events under minimal progress is the
guarded event $\Fm(r)$ defined as follows:
\begin{equation}
  \label{defFm}
  \Fm(r) = \cpl{r}\Rarrow W_{m}
\end{equation}
Definition of the \emph{termination\/} relation under minimal progress
is given by all ordered pairs $a\mapsto b$ satisfying $a\subseteq
\pre(\Fm(b)\open)$: 
\begin{equation}
  \label{termM}
  \Tm=\{\,a\mapsto b\,|\,a\subseteq u\And b\subseteq u\And
  a\subseteq \fix(\Fm(b))\,\}
\end{equation}

\subsubsection{Reachability under MP}\label{reachMP}
The basic relation under minimal progress contains all
ordered pairs $a\mapsto b$ from which we can derive a 
property $x\in a \gg_{m} x\in b$ (\ref{mp}):       
\begin{eqnarray}
\label{defEm}
\lefteqn{\Em=\{\,a\mapsto b\,|\,a\subseteq u\And b\subseteq u\And
  a\cap \cpl{b}\subseteq S(b)\cap \grs(S)\,\}} 
\end{eqnarray} 

From definitions of $\Em$ and $W_{m}$, the proof of premise (a)
of theorem \ref{Tsoundcomp} follows for the case of minimal progress 
$a\mapsto b\in \Em\Imp a\cap \cpl{b}\subseteq W_{m}(b)$: 
\begin{PreuveL}
1 & a\mapsto b\in \Em &  premise\\
2 & a\cap \cpl{b}\subseteq \grs(S)\cap S(b) &  1 and def. (\ref{defEm})\\
3 & a\cap \cpl{b}\subseteq (\grs(S)\; |\; S)(b) &  2 and set transf.\\
4 & a\cap \cpl{b}\subseteq W_{m}(b) &  3 and def. (\ref{Wmp})
\end{PreuveL}

From definition of $\Em$, the implication $a\subseteq b\Imp a\mapsto
b\in \Em$ follows immediately because $a\cap\cpl{b}=\ets$. It proves
premise (b) of theorem \ref{Tsoundcomp} for the case of minimal
progress. 

Now, we use an induction scheme to define the \emph{reachability\/}
relation under minimal progress $\Lm$ similar to definition
\ref{reach}. Therefore $\Lm$ is the smallest relation containing the
base relation $\Em$ and it is both, transitive and disjunctive. 

Finally we prove that the weakest precondition $W_{m}(r)$, for any
$r\in\Pow(u)$ leads to $r$: $W_{m}(r)\mapsto r\in \Lm$ 
\begin{PreuveL}
1 & \grs(S)\cap S(r)\cap \cpl{r}\subseteq \grs(S)\cap S(r) &  trivial\\
2 & \grs(S)\cap S(r)\mapsto r\in \Em &  1 and def. $\Em$\\
3 & W_{m}(r)\mapsto r\in \Em &  2 and (\ref{Wmp})\\
4 & W_{m}(r)\mapsto r\in \Lm &  3 and def. $\Lm$
\end{PreuveL}
This proves premise (c) of theorem \ref{Tsoundcomp} for the case of
minimal progress.

At this time, monotonicity and strictness of $W_m$ and premises (a),
(b) and (c) of theorem \ref{Tsoundcomp} instantiated to the case of
minimal progress have been proved.  Therefore the equality between 
\emph{termination\/} and \emph{reachability\/} relations is stated:
\begin{equation}
  \label{equalMP}
  \Tm = \Lm
\end{equation}

\subsection{Weak Fairness}\label{weak}

In this subsection, we define the \emph{termination\/} and
\emph{reachability\/} relations for weak fairness assumptions. We
prove that premises of theorem \ref{Tsoundcomp}, instantiated to the
case of weak fairness, are satisfied with these definitions. Therefore
we claim the equality between these relations.

\subsubsection{Termination under WF}\label{termWF}
We use the dovetail operator presented in section \ref{set-trans} to
model a \emph{fair loop} for a certain event  $G$ in $\S$: 
\begin{eqnarray}\label{defY}
\lefteqn{Y(q)(G)=\cpl{q}\Rarrow ((S\sco Y(q)(G))\dtl (\grs(G)\;|\;G))}
\end{eqnarray}
The guard $\cpl{q}$ of this loop prevents iteration of the fair choice
in any state belonging to $q$. Informally, we expect that any
execution of $Y(G)(q)$ in any state in $q \cup \grs(G)$
terminates. Execution of $Y(q)(G)$ in $\cpl{q}\cap\grs(G)$ cannot
loops forever because the dovetail operator prevents unlimited
execution of the branch $S\sco Y(q)(G)$. Moreover the set transformer
$\grs(G)\;|\;G$  is always enabled ($\grs(\grs(G)\;|\;G)=u$) and
therefore it will be eventually executed. All our claims are formally
justified by the calculi of termination set and the liberal weakest
precondition of $Y(q)(G)$, for any $q$ and $r$, $r\not=u$ in
$\Pow(u)$:
\begin{eqnarray}
  \label{preY}
  \lefteqn{\pre(Y(q)(G)) = \fix(\cpl{q}\cap G(\ets)\Rarrow
    \cpl{S(q)}\; |\; S)}\\
  \label{LiY}
  \lefteqn{\Li(Y(q)(G))(r)=\FIX(\cpl{q}\Rarrow (\grs(G)\cap
    G(r)\;|\;S))} 
\end{eqnarray}
These calculi follow from definitions of set transformers given in
section \ref{set-trans} and the extreme solutions of the recursive
equations generated. The proof of (\ref{preY}) and (\ref{LiY}) is
given in  appendix \ref{AppendWeak}. Moreover, in appendix \ref{AppendWeak}
appears the proof of  the following inclusion, for any $q$ and $r$ in $\Pow(u)$:
\begin{equation}
\Li(Y(q)(G))(r)\subseteq \pre(Y(q)(G)) \label{libInpre}
\end{equation}
(\ref{libInpre}), and the pairing condition, give us the set
transformer associated with the fair loop:
\begin{equation}
  \label{strY}
  Y(q)(G)(r)=\FIX(\cpl{q}\Rarrow (\grs(G)\cap G(r)\; |\; S))
\end{equation}
From this definition  follows the monotonicity of $Y(q)(G)$, which is
proved in appendix \ref{AppendWeak}.

The fair loop $Y(q)(G)$ models a \emph{fair $G$-step\/} in the iteration
of events under weak fairness assumptions. We say that $G$ is the
helpful event in this $G$-step. A \emph{fair step} in the iteration of
events is  modeled by the following set transformer:
\begin{equation}
  \label{Wwf}
  W_{w} = \lambda r\cdot (r\subseteq u\;|\;\bigcup G\cdot
  (G\in\S\;|\;Y(r)(G)(r))) 
\end{equation}
From (\ref{strY}) follows $\grs(Y(q)(G))=\cpl{q}$ for any $G$ in $\S$
and $q\in\Pow(u)$, therefore  the strictness of $W_w$ follows. On the
other hand, from monotonicity of $Y(q)(G)$ follows the monotonicity of
$W_w$. These three proofs are given in appendix \ref{AppendWeak}.

The body of the iteration of events under weak fairness is the
guarded event $\Fw(r)$ defined as follows:
\begin{equation}
  \label{defFW}
  \Fw(r) = \cpl{r}\Rarrow W_{w}
\end{equation}
Definition of the \emph{termination\/} relation under weak fairness
is:
\begin{equation}
  \label{termW}
  \Tw=\{\,a\mapsto b\,|\,a\subseteq u\And b\subseteq u\And
  a\subseteq \fix(\Fw(b))\,\}
\end{equation}

\subsubsection{Reachability under WF}\label{reachWF}
We define the basic relation  $\Ee(G)$ for a helpful event $G$, 
as the set of pairs $a\mapsto b$ from which we can derive a property 
$G\cdot x\in a \gg_{w} x\in b$ (\ref{wf}): 
\begin{eqnarray}\label{defE'}
\lefteqn{\Ew(G)=\{\,a\mapsto b\,|\,a\subseteq u\And b\subseteq u\And
  a\cap \cpl{b}\subseteq S(a\cup b)\cap \cpl{G(\ets)}\cap G(b)\,\}} 
\end{eqnarray}
Now, the basic relation for weak fairness is:
\begin{eqnarray}\label{defEw}
\lefteqn{\Ew=\bigcup G\cdot (G\in \S\;|\;\Ew'(G))}
\end{eqnarray}

The proof of premise (a) of theorem \ref{Tsoundcomp} instantiated to
weak fairness requires the following property which is proved in
appendix \ref{AppendWeak}:
\begin{equation}
  \label{propE'}
  \forall G\cdot (G\in\S \And a\mapsto b\in \Ew'(G)\Imp a\subseteq Y(b)(G)(b))
\end{equation}
Using (\ref{propE'}), the proof of $a\mapsto b\in \Ew\Imp a\cap
\cpl{b}\subseteq W_{w}(b)$ is:  
\begin{PreuveL}
1 & \forall G\cdot (G\in \S\Imp (a\mapsto b\in \Ew'(G)\Imp a\subseteq
Y(b)(G)(b))) &  (\ref{propE'})\\ 
2 & \exists G\cdot (G\in \S\And a\mapsto b\in \Ew'(G))\Imp a\subseteq
W_{w}(b) &  1 and (\ref{Wwf}). \\ 
3 & a\mapsto b\in \bigcup G\cdot (G\in \S\;|\;\Ew'(G))\Imp a\subseteq W_{w}(b) &  2\\
4 & a\mapsto b\in \Ew\Imp a\subseteq W_{w}(b) &  3 and (\ref{defEw})\\
5 & a\mapsto b\in \Ew\Imp a\cap \cpl{b}\subseteq W_{w}(b) &  4 and
$b\subseteq W_{w}(b)$  
\end{PreuveL}

From (\ref{defE'}) immediately follows $a\mapsto b\in\Ew'(G)$, for any
$G$ in $\S$ if $a\subseteq b$ holds, and from (\ref{defEw})
follows $a\subseteq b \Imp a\mapsto b\in\Ew$. This proves premise (b) of
theorem \ref{Tsoundcomp}.

We use an induction scheme to define the \emph{reachability\/}
relation under weak fairness $\Lw$ similar to definition
\ref{reach}. Therefore $\Lw$ is the smallest relation containing the
base relation $\Ew$ and it is both, transitive and disjunctive. 

From (\ref{strY}) and (\ref{defE'}) follows the property:
\begin{eqnarray}
\label{propEwp}
\lefteqn{\forall (G,r)\cdot (G\in \S\And r\subseteq u\Imp
  Y(r)(G)(r)\mapsto r\in \Ew'(G))} 
\end{eqnarray}
We use this property to prove the premise (c) of theorem
\ref{Tsoundcomp}: $W_{w}(r)\mapsto r\in \Lw$ as follows:
\begin{PreuveL}
1 & \forall G\cdot (G\in \S\Imp Y(r)(G)(r)\mapsto r\in \Ew'(G)) &
from (\ref{propEwp}) \\ 
2 & \forall G\cdot (G\in \S\Imp \Ew'(G)\subseteq \Ew) &  def. $\Ew$\\
3 & \forall G\cdot (G\in \S\Imp Y(r)(G)(r)\mapsto r\in \Ew) &  2 and 1\\
4 & \forall G\cdot (G\in \S\Imp Y(r)(G)(r)\mapsto r\in \Lw) &  def. $\Lw$\\
5 & \{\,Y(r)(G)(r)\,|\,G\in \S\,\}\times \{r\}\subseteq \Lw &  4\\
6 & \bigcup(\{\,Y(r)(G)(r)\,|\,G\in \S\,\})\mapsto r\in \Lw &  5 and SDR\\
7 & \bigcup G\cdot (G\in \S\;|\;Y(r)(G)(r))\mapsto r\in \Lw &  6\\
8 & W_{w}(r)\mapsto r\in \Lw &  7 and def. $F$
\end{PreuveL}

At this time \emph{termination\/} ($\Tw$), basic relation
($\Ew$) and \emph{reachability\/} ($\Lm$) relations for weak fairness
assumptions have been defined. Monotonicity and strictness of the set
transformer $W_{w}$, and premises (a), (b) and (c) of theorem
\ref{Tsoundcomp} instantiated to the case of weak fairness have been
proved.  Therefore, the equality between \emph{termination\/} and
\emph{reachability\/} relations under weak fairness is stated:
\begin{equation}\label{equalWF}
  \Tw = \Lw
\end{equation}

\section{Deriving Liveness Properties}\label{deriving}
In this section we present two examples where we show practical
usefulness of equalities between \emph{termination\/} and
\emph{reachability\/} relations under minimal progress and weak
fairness assumptions. This section is divided in three parts. In the
first part we state and prove the \emph{Variant Theorem\/}, which
allows us to prove termination of iterations over a set transformer if
a variant decreases. In the second part we use this theorem to prove a
sufficient condition allowing derivation of liveness properties under
minimal progress.  Finally, we give another sufficient condition to
derive a liveness property under minimal progress when a similar
property holds in weak fairness assumptions

\subsection{The Variant Theorem}\label{svarth}
The variant theorem allows us to prove termination of iteration of
conjunctive set transformers. This theorem considers a total function
which maps each element of the state space to an element of a well
founded order and a set which is invariant at each iteration of the
set transformer. The theorem states that if any execution of the set
transformer starting in a state in the invariant set and a certain
value of the variant function, terminates in a state where the value
of the variant is decremented, then the invariant set is contained in
the termination set of the iteration of the set transformer. Formally,
the theorem is stated as follows:
%
\begin{theorem}\label{varth}
  \emph{(Variant Theorem)}\\
  Let $V\in u\Totfunc \Nat$, $v=\lambda n\cdot (n\in
  \Nat\;|\;\{\,z\,|\,z\in u\And V(z)=n\,\})$ and $\m{v'}=\lambda
  n\cdot (n\in \Nat\;|\;\{\,z\,|\,z\in u\And V(z)< n\,\})$. For any
  conjunctive set transformer $f$ in $\Pow(u)\Totfunc \Pow(u)$ and $p$
  in $\Pow(u)$, such that $v(n)\cap p\subseteq f(\m{v'}(n))$ and
  $p\subseteq f(p)$ , for any $n$ in $\Nat$, the following inclusion
  holds:
\[ p\subseteq \fix(f) \]
\end{theorem}
%
The proof of this theorem uses the following equalities:
\begin{eqnarray}
\label{varthA}
\lefteqn{\forall n\cdot (n\in \Nat\Imp \m{v'}(n)=\bigcup i\cdot (i\in \Nat\And i< n\;|\;v(i)))}\\
\label{varthB}
\lefteqn{\bigcup i\cdot (i\in \Nat\;|\;\m{v'}(i+1))=\bigcup i\cdot (i\in \Nat\;|\;v(i))}\\
\label{varthC}
\lefteqn{\bigcup i\cdot (i\in \Nat\;|\;v(i))=u}
\end{eqnarray}
and the following property:
\begin{eqnarray}
\label{varth2} 
&&\forall n\cdot (n\in \Nat\Imp \bigcup i\cdot (i\in\Nat\And i\leq
n\;|\;v(i))\cap p\subseteq f^{n+1}) 
\end{eqnarray}
which are proved, under the assumptions of the theorem \ref{varth}, in appendix
\ref{AppendDeriving}. The proof of theorem \ref{varth} is as follows:
\begin{PreuveL}
1 & \forall n\cdot (n\in \Nat\Imp \bigcup i\cdot (i\in \Nat\And i\leq n\;|\;v(i))\cap p\subseteq \fix(f)) &  (\ref{varth2}) and (\ref{Pfix1})\\
2 & \forall n\cdot (n\in \Nat\Imp \m{v'}(n+1)\cap p\subseteq \fix(f)) &  from (\ref{varthA}) and 1\\
3 & \bigcup i\cdot (i\in \Nat\;|\;\m{v'}(i+1))\cap p\subseteq \fix(f) &  2\\
4 & \bigcup i\cdot (i\in \Nat\;|\;v(i))\cap p\subseteq \fix(f) &  3 and (\ref{varthB})\\
5 & u\cap p\subseteq \fix(f) &  4 and (\ref{varthC})\\
6 & p\subseteq \fix(p) &  5 and $p\subseteq u$  
\end{PreuveL}

\subsection{A Sufficient Condition for Minimal
  Progress}\label{sufmp}
A system reaches a certain set from any set of starting states 
under minimal progress, if the set of depart is invariant in the
system, it is contained in the guard of the system and
each execution of the system decrements a variant. Formally, these
conditions are stated as follows:
\begin{center}
\begin{tabular}{|l|l|}\hline
 \ ~~~~~~~~~ANTECEDENT &\ ~CONSEQUENT~ \\ \hline
 & \\[-2mm]
\ $\forall n\cdot (n\in \Nat\Imp a\cap \cpl{b}\cap v(n)\subseteq
S(\m{v'}(n)))$  &\ $a\mapsto b\in \Lm$  \\[1mm] 
\ $a\cap \cpl{b}\subseteq \grs(S)\cap S(a)$  & \\[1mm]
\hline
\end{tabular}
\end{center}
We remark from definition (\ref{defFm}), that $\Fm(b)$ is a
conjunctive set transformer. From this remark the proof of the rule is
as follows:
\begin{PreuveL}
1 & \forall n\cdot (n\in \Nat\Imp a\cap \cpl{b}\cap v(n)\subseteq S(\m{v'}(n))\cap \grs(S)) &  from premises\\
2 & \forall n\cdot (n\in \Nat\Imp a\cap v(n)\subseteq \Fm(b)(\m{v'}(n))) &  1 and (\ref{defFm})\\
3 & a\subseteq \Fm(b)(a) &  premise and (\ref{defFm})\\
4 & a\subseteq \fix(\Fm(b)) &  3, 2, theorem \ref{varth}\\
5 & a\mapsto b\in \Tm &  4 and (\ref{termM})\\
6 & a\mapsto b\in \Lm &  5, equality (\ref{equalMP})
\end{PreuveL}

Antecedent of this rule corresponds to sufficient conditions in
\cite{AbrMusIDCB} to prove liveness properties and it is the only rule
concerning the proof of liveness properties. Soundness of this rule is
proved here in a more direct way. 

Soundness of this rule is given without reasoning over state-traces,
taking advantage of the fixpoint semantics approach.

\subsection{From Weak Fairness to Minimal Progress}\label{wfmp}
\newcommand{\Ygb}{Y_{B}^{G}}
\newcommand{\Ww}{W_{w}}
\newcommand{\Fwb}{F}
\newcommand{\Fi}{F^{\a}}
\newcommand{\FF}{F}

Using the variant theorem, we prove a sufficient condition to establish
that a liveness property under minimal progress, follows from a
corresponding property proved under weak fairness and from the
decrement of a variant:
\begin{center}
\begin{tabular}{|l|l|}\hline
 \ ~~~~~~~~~ANTECEDENT &\ ~CONSEQUENT~ \\ \hline
 & \\[-2mm]
\ $\forall n\cdot (n\in \Nat\Imp \cpl{b}\cap v(n)\subseteq
S(\m{v'}(n)))$  &\ $a\mapsto b\in \Lm$  \\[1mm] 
\ $a\mapsto b\in\Lw$  & \\[1mm]
\hline
\end{tabular}
\end{center}
The proof of these conditions is given by the \emph{Variant Theorem}.
In order to apply the theorem, we need to identify an
invariant set under $\Fm(b)$. However, as the sets $a$
and $b$ cannot be proved as invariants, we prove that the least
fixpoint of $\Fw(b)$ is invariant under $\Fm(b)$, that is
$\fix(\Fw(b))\subseteq \Fm(b)(\fix(\Fw(b)))$. This proof requires the
following lemma:
\begin{eqnarray}
\label{wfmp1}
\lefteqn{\forall \a\cdot (\Fw(b)^{\a}\subseteq b\cup (\grs(S)\cap
  S(\fix(\Fw(b))))} 
\end{eqnarray}

\renewcommand{\Fwb}{\Fw(b)}
\noindent The proof of (\ref{wfmp1}) is done by transfinite
induction; it is presented in appendix \ref{AppendDeriving}.
Using (\ref{wfmp1}), the proof of sufficient conditions are as
follows:\\
\begin{PreuveL}
1 & \fix(\Fwb)\subseteq b\cup (\grs(S)\cap S(\fix(\Fwb))) &  Theorem \ref{tglim}, (\ref{wfmp1})\\
2 & \fix(\Fwb)\subseteq \Fm(b)(\fix(\Fwb)) &  1 and (\ref{defFm})\\
3 & \forall n\cdot (n\in \Nat\Imp \fix(\Fwb)\cap v(n)\subseteq b\cup \grs(S)\cap S(\m{v'}(n))) &  2 and premise\\
4 & \forall n\cdot (n\in \Nat\Imp \fix(\Fwb)\cap v(n)\subseteq \Fm(b)(\m{v'}(n))) &  3 and (\ref{defFm})\\
5 & \fix(\Fwb)\subseteq \fix(\Fm(b)) &  4,2 and th. \ref{varth}\\
6 & a\mapsto b\in \Tw &  premise, eq. (\ref{equalWF})\\
7 & a\subseteq \fix(\Fwb) &  6 and def. $\Tw$\\
8 & a\subseteq \fix(\Fm(b)) &  7 and 5\\
9 & a\mapsto b\in \Lm &  8, def. $\Tm$, eq. (\ref{equalMP})
\end{PreuveL}

\section{Conclusions}\label{conc}

We have presented a fixpoint semantics of event systems under minimal
progress and weak fairness assumptions. Then we have proved soundness
and completeness of rules for deriving \emph{leads-to\/} properties
under weak fairness and minimal progress assumptions. Finally we
have proved sufficient conditions to guarantee a liveness property under
minimal progress in two cases of hypothesis: every event decrements a
variant under an invariant, or every event decrements a variant and the
property holds under weak fairness.

The development of our semantics is structured. First a general
framework is established without concerns of fairness, and our notions
of termination and reachability are elaborated. Soundness and
completeness of rules for \emph{leads-to\/} are proved in this
framework. The general framework is then instantiated to the cases of
minimal progress and weak fairness assumptions and the corresponding
results are proved. Each element in our models has a concrete
representation as a set transformer. In particular, we stress how the
weak fairness assumption is modeled by the dovetail operator.

We have stated a simple form of the variant theorem and  given a simple
proof of it. We remark the usefulness of this theorem in the proofs of
liveness properties. Particularly we note the importance of conditions
which guarantee the derivation of a certain liveness property ${\cal
  P}$ under minimal progress if ${\cal P}$ holds under weak fairness,
and every element of the system decrements a variant. This is a new
result which gives the possibility to \emph{implement\/} fairness in a
system. 

As a future work we investigate how our approach can be managed to
deal with refinement of event systems. Another line will be to consider how
to instantiate the general framework for strong fairness.

\bibliographystyle{plain} \bibliography{./docs1}

\appendix
\newpage
\mbox{}
\vfill
\centerline{\bf \large ANNEXES }
\vfill
\newpage
\section{\emph{leads-to} as Relation Between Predicates or Sets} 
\label{equsetpred}
In order to guarantee that \emph{leads-to\/} ($\leadsto$), as a relation
between predicates on the system state, and the relation $\L$ between
subsets of $u$ are equivalent, we supposed the following equivalence:
\[
\parbox{.937\textwidth}{$P(x)\leadsto Q(x)\equiv \{\,z\,|\,z\in u\And P(z)\,\}\mapsto
\{\,z\,|\,z\in u\And Q(z)\,\}\in \L$\hfill (\ref{defequlto})} 
\]
In the following paragraphs we give the proof of this equivalence. It 
is founded in the fact that \emph{leads-to\/} relation of {\sc unity}
logic, as pointed in \cite{RaJuMDPR}, can be defined by an induction
scheme, similar to definition \ref{reach}. That is, $\leadsto$ as a
relation between predicates, $\leadsto \; \subseteq \Pred\times\Pred$,
where \Pred\ is the set of predicates on the space of the state
variable $x$, is the smallest relation satisfying rules BRL, TRA and
DSJ given in section \ref{general}\footnote{The infix notation
  $P\leadsto Q$ is used to state that $P\mapsto Q\in\;\leadsto$, for
  any predicate $P$ and $Q$ in \Pred.}:
\begin{description}
\item[BRL:] $E\subseteq\; \leadsto\;$
\item[TRA:] $\leadsto^{2}\;\subseteq\; \leadsto$
\item[DSJ:] $\forall m\cdot (m\in M\Imp P(m)\leadsto Q) \Imp\exists
  m\cdot (m\in M\And P(m)) \leadsto Q$ 
\end{description}
where $E$ is the set of couples of predicates satisfying the
\emph{ensures\/} relation:
\begin{equation}
\label{predensures}
E=\{\,P\mapsto Q\,|\,P\gg Q\,\}
\end{equation}

We recall that $\gg$ relation must be instantiated to $\gg_m$ relation
under minimal progress hypothesis  or $\gg_w$ relation under weak
fairness assumptions, in similar way to the instantiation of
the basic relation $\E$. In order to give the proof of equivalence
(\ref{defequlto}), at this time we suppose that $E$ and $\E$ satisfy
the following properties:
\begin{eqnarray}
\label{equensures1}
\lefteqn{\forall (P,Q)\cdot (P\in \Pred\And Q\in \Pred\Imp\nonumber}\\&&~~~~~~P(x)\gg Q(x)\equiv \{\,z\,|\,z\in u\And P(z)\,\}\mapsto \{\,z\,|\,z\in u\And Q(z)\,\}\in \E)\\
\label{equensures2}
\lefteqn{\forall (p,q)\cdot (p\subseteq u\And q\subseteq u\Imp p\mapsto q\in \E\equiv x\in p\gg x\in q)}
\end{eqnarray}
We give below the proof of these properties, instantiated to weak
fairness or minimal progress assumptions.

The proof of (\ref{defequlto}) is given in two parts:
\begin{eqnarray}
\label{equlto1}
\lefteqn{P(x)\leadsto Q(x)\Imp \{\,z\,|\,z\in u\And P(z)\,\}\mapsto \{\,z\,|\,z\in u\And Q(z)\,\}\in \L}\\
\label{equlto2}
\lefteqn{\{\,z\,|\,z\in u\And P(z)\,\}\mapsto \{\,z\,|\,z\in u\And Q(z)\,\}\in \L\Imp P(x)\leadsto Q(x)}
\end{eqnarray}

\subsubsection*{Proof of (\ref{equlto1})}\mbox{}\\
\noindent Let $\P$ be the following set:
\begin{eqnarray*}
\lefteqn{\P=\{\,P\mapsto Q\,|\,(P\leadsto Q) \And
  \{\,z\,|\,z\in u\And P(z)\,\}\mapsto \{\,z\,|\,z\in u\And
  Q(z)\,\}\in \L\,\}} 
\end{eqnarray*}
From this definition follows $\P\subseteq\;\leadsto$. Inclusion
$\leadsto\;\subseteq \P$ is proved below by structural induction. From
these inclusions, follows the equality $\P =\; \leadsto$. Finally,
from this equality follows (\ref{equlto1}):
\bop
\begin{Preuve}
\P=\;\leadsto\\
\Com{\equiv}{ $\P=\;\leadsto\cap \;\{\,P\mapsto Q\,|\,\{\,z\,|\,z\in u\And P(z)\,\}\mapsto \{\,z\,|\,z\in u\And Q(z)\,\}\in \L\,\}$ }\\
\leadsto\;\subseteq \{\,P\mapsto Q\,|\,\{\,z\,|\,z\in u\And P(z)\,\}\mapsto \{\,z\,|\,z\in u\And Q(z)\,\}\in \L\,\}\\
\Com{\equiv}{ }\\
\forall (P,Q)\cdot (P\leadsto Q\Imp \{\,z\,|\,z\in u\And P(z)\,\}\mapsto \{\,z\,|\,z\in u\And Q(z)\,\}\in \L)
\end{Preuve}
\eop

In order to proof $\leadsto\;\subseteq \P$, the following proofs are required:
\begin{itemize}
\item $E\subseteq \P$.
\item $\P^{2}\subseteq \P$.
\item $\forall i\cdot (i\in I\Imp P(i)\mapsto Q\in \P)\Imp \exists
  i\cdot (i\in I\And P(i)) \mapsto Q \in \P$
\end{itemize}

\noindent{\bf Proof of Base Case}
\begin{PreuveL}
1 & P\gg Q\Imp \{\,z\,|\,z\in u\And P(z)\,\}\mapsto \{\,z\,|\,z\in u\And Q(z)\,\}\in \E &  from (\ref{equensures1})\\
2 & P\gg Q\Imp \{\,z\,|\,z\in u\And P(z)\,\}\mapsto \{\,z\,|\,z\in u\And Q(z)\,\}\in \L &  1 and def. \ref{reach}\\
3 & P\gg Q\Imp P\leadsto Q &  BRA\\
4 & P\mapsto Q\in E\Imp P\mapsto Q\in \P &  3, 2 and (\ref{predensures})\\
5 & E\subseteq \P &  4
\end{PreuveL}
\eop

\noindent{\bf Proof of Transitivity}\\
It follows from $P\mapsto Q\in \P\And Q\mapsto R\in \P\Imp P\mapsto
R\in \P$:
\begin{PreuveL}
1 & P\mapsto Q\in \P\And Q\mapsto r\in \P &  premise\\
2 & P\leadsto R &  1, def. $\P$, TRA\\
3 & \{\,z\,|\,z\in u\And P(z)\,\}\mapsto \{\,z\,|\,z\in u\And
    R(z)\,\}\in \L &  1, def. $\P$, STR\\ 
4 & P\mapsto R\in \P &  3, 2 and def. $\P$
\end{PreuveL}
\eop

\noindent{\bf Proof of Disjunction}
\begin{PreuveCL}{N}
1 & \forall i\cdot (i\in I\Imp P(i)\mapsto Q\in \P) & ; premise\\
2 & \exists i\cdot (i\in I\And P(i))\leadsto Q & ; 1, def. $\P$, DSJ\\
3 & \mbox{$\forall i\cdot (i\in I\Imp \{\,z\,|\,z\in u\And
  P(i)(z)\,\}\mapsto \{\,z\,|\,z\in u\And Q(z)\,\}\in \L)$} &
; 1, def. $\P$\\
4 & \{\,\{\,z\,|\,z\in u\And P(i)(z)\,\}\,|\,i\in I\,\}\times
\{\,z\,|\,z\in u\And Q(z)\,\}\subseteq \L &; 3\\
5 & \mbox{$\bigcup(\{\,\{\,z\,|\,z\in u\And P(i)(z)\,\}\,|\,i\in
  I\,\})\mapsto \{\,z\,|\,z\in u\And Q(z)\,\}\in \L$} &  ;
4, SDJ\\
6 & \mbox{$\{\,z\,|\,z\in u\And \exists i\cdot (i\in I\And
  P(i)(z))\,\}\mapsto \{\,z\,|\,z\in u\And Q(z)\,\}\in \L$} &; 5\\
7 & \exists i\cdot (i\in I\And P(i))\mapsto Q\in \P &;  6, 2 and def. $\P$
\end{PreuveCL}
\eop

\subsubsection*{Proof of (\ref{equlto2})}\mbox{}\\
\noindent This proof is similar to the proof of (\ref{equlto1}). Let
$\Q$ be the following set: 
\begin{eqnarray*}
\lefteqn{\Q=\{\,p\mapsto q\,|\,p\mapsto q\in \L\And x\in p\leadsto x\in q\,\}}
\end{eqnarray*}
From this definition follows $\Q\subseteq \L$. Inclusion
$\L\subseteq\Q$ is proved below by structural induction. From these
inclusions follows equality $\Q = \L$. Now, from this equality
follows inclusion $\L\subseteq \{\,p\mapsto q\,|\,p\mapsto q\in
\L\And x\in p\leadsto x\in q\,\}$:
\begin{Preuve}
\Q=\L\\
\Com{\equiv}{ $Q=\L\cap \{\,p\mapsto q\,|\,p\subseteq u\And q\subseteq u\And x\in p\leadsto x\in q\,\}$ }\\
\L\subseteq \{\,p\mapsto q\,|\,p\subseteq u\And q\subseteq u\And x\in p\leadsto x\in q\,\}
\end{Preuve}
Finally, from this inclusion, and taking $\{\,z\,|\,z\in u\And
P(z)\,\}\mapsto \{\,z\,|\,z\in u\And Q(z)\,\}\in \L$ as a premise, the
conclusion $P\leadsto Q$ of (\ref{equlto2}) follows.\\
\eop

In order to prove $\L\subseteq\Q$ the following proofs are required:
\begin{itemize}
\item $\E\subseteq \Q$.
\item $\Q^{2} \subseteq \Q$.
\item $l\times \{q\}\subseteq \Q\Imp \bigcup(l)\mapsto q\in \Q$ 
\end{itemize}

\noindent{\bf Proof of Base Case}
\begin{PreuveL}
1 & p\mapsto q\in \E\Imp x\in p\gg x\in q &  from (\ref{equensures2})\\
2 & p\mapsto q\in \E\Imp x\in p\leadsto x\in q &  1 and BRL\\
3 & p\mapsto q\in \E\Imp p\mapsto q\in \L &  SBR\\
4 & p\mapsto q\in \E\Imp p\mapsto q\in \Q &  3, 2, def. $\Q$\\
5 & \E\subseteq \Q &  4
\end{PreuveL}
\eop

\noindent{\bf Proof of Transitivity}
\begin{PreuveL}
1 & p\mapsto q\in \Q\And q\mapsto r\in \Q &  premise\\
2 & p\mapsto q\in \L\And q\mapsto r\in \L &  1, def $\Q$\\
3 & (x\in p\leadsto x\in q)\And (x\in q\leadsto x\in r) &  1, def $Q$\\
4 & p\mapsto r\in \L\And x\in p\leadsto x\in r &  2, 3, TRA,STR\\
5 & p\mapsto r\in \Q &  4, def. $\Q$
\end{PreuveL}
\eop

\noindent{\bf Proof of Disjunction}
\begin{PreuveL}
1 & l\times \{q\}\subseteq \Q &  premise\\
2 & l\times \{q\}\subseteq \L &  1, def. $\Q$\\
3 & \bigcup(l)\mapsto q\in \L &  2, SDR\\
4 & l\times \{q\}\subseteq \{\,a\mapsto b\,|\,a\subseteq u\And b\subseteq u\And x\in a\leadsto x\in b\,\} &  1, def. $\Q$\\
5 & \forall s\cdot (s\in l\Imp x\in s\leadsto x\in q) &  $l\subseteq\Pow(u)$, 4\\
6 & \exists s\cdot (s\in l\And x\in s)\leadsto x\in q &  5, DSJ\\
7 & x\in \bigcup(l)\leadsto x\in q &  6\\
8 & \bigcup(l)\mapsto q\in \Q &  7, 3, def. $\Q$
\end{PreuveL}
\eop

\subsection{Instantiation of \emph{ensures\/} to Minimal Progress}
In this section, properties (\ref{equensures1}) and (\ref{equensures2}) are
proved when the \emph{ensures\/} relation is instantiated to a minimal
progress assumption.

Definition of \emph{ensures\/} relation under minimal progress
assumptions ($\gg_m$), without considering the strongest invariant, is
given by the following definition:
\begin{eqnarray*}
\lefteqn{\forall (P,Q)\cdot (P\in \Pred\And Q\in \Pred\Imp P(x)\gg_{m} Q(x)\equiv }\\
&&\forall x\cdot (I(x)\And P(x)\And \neg Q(x)\Imp \grd(\sub(S))\And [\sub(S)]\,Q(x)))
\end{eqnarray*}
where $\sub(S)$ denotes the \emph{generalized substitution\/}
associated with set transformer $S$, and for any generalized
substitution $T$, the predicate $\grd(T)$ is equivalent to $\neg [T]\;\False$.

\noindent{\bf Proof of (\ref{equensures1})}
\begin{Preuve}
P(x)\gg_{m} Q(x)\\
\Com{\equiv}{ def. $\gg_m$ }\\
\forall x\cdot (I(x)\And P(x)\And \neg Q(x)\Imp \grd(\sub(S))\And [\sub(S)]\,Q(x))\\
\Com{\equiv}{ set theory }\\
\forall x\cdot (x\in \{\,z\,|\,I(z)\And P(z)\,\}\cap \{\,z\,|\,I(z)\And \neg Q(x)\,\}\Imp I(x)\And \grd(\sub(S))\And [\sub(S)]\,Q(x))\\
\Com{\equiv}{ $I(x)\Imp [\sub(S)]\,I(x)$, conjunctive $S$ }\\
\mbox{$\forall x\cdot (x\!\in \!\{\,z\,|\,I(z)\!\And\! P(z)\,\}\!\cap\! \{\,z\,|\,I(z)\!\And \!\neg Q(x)\,\}\!\Imp\! I(x)\!\And\! \grd(\sub(S))\!\And\! [\sub(S)]\,I(x)\!\And \!Q(x))$}\\
\Com{\equiv}{ set transformers }\\
\forall x\cdot (x\in \{\,z\,|\,I(z)\And P(z)\,\}\cap \{\,z\,|\,I(z)\And \neg Q(x)\,\}\Imp x\in \grs(S)\cap S(\{\,z\,|\,I(z)\And Q(z)\,\}))\\
\Com{\equiv}{ set theory, $u=\{\,z\,|\,I(z)\,\}$ }\\
\{\,z\,|\,z\in u\And P(z)\,\}\cap \cpl{\{\,z\,|\,z\in u\And Q(x)\,\}}\subseteq \grs(S)\cap S(\{\,z\,|\,z\in u\And Q(z)\,\})\\
\Com{\equiv}{ def. $\Em$ }\\
\{\,z\,|\,z\in u\And P(z)\,\}\mapsto \{\,z\,|\,z\in u\And Q(x)\,\}\in \Em
\end{Preuve}
\eop

\noindent{\bf Proof of (\ref{equensures2})}
\begin{Preuve}
p\mapsto q\in \Em\\
\Com{\equiv}{ def. $\Em$ }\\
p\cap \cpl{q}\subseteq S(q)\cap \grs(S)\\
\Com{\equiv}{ }\\
\forall x\cdot (x\in p\cap \cpl{q}\Imp x\in S(q)\cap \grs(S))\\
\Com{\equiv}{ $x\in p\Imp I(x)$ }\\
\forall x\cdot (I(x)\And x\in p\And \neg x\in q\Imp x\in S(q)\cap \grs(S))\\
\Com{\equiv}{ set transformers }\\
\forall x\cdot (I(x)\And x\in p\And \neg x\in q\Imp [\sub(S)]\,x\in q\And \grd(S))\\
\Com{\equiv}{ def. $\gg_m$ }\\
x\in p\gg_{m} x\in q
\end{Preuve}
\eop

\subsection{Instantiation of \emph{ensures\/} to Weak Fairness}
In this section, properties (\ref{equensures1}) and (\ref{equensures2}) are
proved when the \emph{ensures\/} relation ($\gg$) is instantiated to a weak
fairness assumption relation ($\gg_w$). The instantiation under this
condition is given by the following equivalence:
\begin{eqnarray*}
\lefteqn{P(x)\gg Q(x)\equiv \exists G\cdot (G\in \S\And G\cdot P(x)\gg_{w} Q(x))}
\end{eqnarray*}
Definition of \emph{ensures\/} relation under weak fairness
assumptions ($\gg_m$), without considering the strongest invariant, is
given by the following definition:
\begin{eqnarray*}
\lefteqn{\forall (P,Q)\cdot (P\in \Pred\And Q\in \Pred\Imp G\cdot P(x)\gg_{w} Q(x)\equiv }\\
&&\forall x\cdot (I(x)\And P(x)\And \neg Q(x)\Imp ([\sub(S)]\,(P(x)\Or
Q(x)))\And \grd(\sub(G))\And [\sub(G)]\,Q(x))) 
\end{eqnarray*}
where $\sub(S)$ and $\sub(G)$ denote the \emph{generalized substitutions\/}
associated with  set transformers $S$ and $G$.

The proof follows from the following equivalences which are proved below:
\begin{eqnarray}
\label{equensures1a}
\lefteqn{G\cdot P(x)\gg_{w} Q(x)\equiv \{\,z\,|\,z\in u\And P(z)\,\}\mapsto \{\,z\,|\,z\in u\And Q(x)\,\}\in \Ee(G)}\\
\label{equensures2a}
\lefteqn{p\mapsto q\in \Ee(G)\equiv G\cdot x\in p\gg_{w} x\in q}
\end{eqnarray}

\noindent{\bf Proof of (\ref{equensures1})}\\
Implication from left to right follows from (\ref{equensures1a})
\begin{Preuve}
G\cdot P(x)\gg_{w} Q(x)\equiv \{\,z\,|\,z\in u\And P(z)\,\}\mapsto \{\,z\,|\,z\in u\And Q(x)\,\}\in \Ee(G)\\
\Com{\Imp}{ $\Ew=\bigcup G\cdot (G\in \S\;|\;\Ee(G))$ }\\
G\cdot P(x)\gg_{w} Q(x)\Imp \{\,z\,|\,z\in u\And P(z)\,\}\mapsto \{\,z\,|\,z\in u\And Q(x)\,\}\in \Ew\\
\Com{\Imp}{  }\\
\exists G\cdot (G\in \S\And G\cdot P(x)\gg_{w} Q(x))\Imp \{\,z\,|\,z\in u\And P(z)\,\}\mapsto \{\,z\,|\,z\in u\And Q(x)\,\}\in \Ew
\end{Preuve}

\noindent Implication from right to left follows from
(\ref{equensures1a}) \\
\begin{Preuve}
G\cdot P(x)\gg_{w} Q(x)\equiv \{\,z\,|\,z\in u\And P(z)\,\}\mapsto \{\,z\,|\,z\in u\And Q(x)\,\}\in \Ee(G)\\
\Com{\Imp}{ }\\
\{\,z\,|\,z\in u\And P(z)\,\}\mapsto \{\,z\,|\,z\in u\And Q(x)\,\}\in \Ee(G)\Imp \exists G\cdot (G\in \S\And G\cdot P(x)\gg_{w} Q(x))\\
\Com{\Imp}{ }\\
\exists G\cdot (G\in \S\And \{\,z\,|\,z\in u\And P(z)\,\}\mapsto \{\,z\,|\,z\in u\And Q(x)\,\}\in \Ee(G))\Imp \exists G\cdot (G\in \S\And G\cdot P(x)\gg_{w} Q(x))\\
\Com{\Imp}{ }\\
\{\,z\,|\,z\in u\And P(z)\,\}\mapsto \{\,z\,|\,z\in u\And Q(x)\,\}\in \bigcup G\cdot (G\in \S\;|\;\Ee(G))\Imp \exists G\cdot (G\in \S\And G\cdot P(x)\gg_{w} Q(x))\\
\Com{\equiv}{ def. $\Ee$ }\\
\{\,z\,|\,z\in u\And P(z)\,\}\mapsto \{\,z\,|\,z\in u\And Q(x)\,\}\in \Ew\Imp \exists G\cdot (G\in \S\And G\cdot P(x)\gg_{w} Q(x))
\end{Preuve}
\eop

\noindent{\bf Proof of (\ref{equensures2})}\\
Implication from left to right follows from (\ref{equensures2a})
\begin{Preuve}
p\mapsto q\in \Ee(G)\equiv G\cdot x\in p\gg_{w} x\in q\\
\Com{\Imp}{ $\Ew=\bigcup G\cdot (G\in \S\;|\;\Ee(G))$ }\\
p\mapsto q\in \Ee(G)\Imp \exists G\cdot (G\in \S\And G\cdot x\in p\gg_{w} x\in q)\\
\Com{\Imp}{ }\\
\exists G\cdot (G\in \S\And p\mapsto q\in \Ee(G))\Imp \exists G\cdot (G\in \S\And G\cdot x\in p\gg_{w} x\in q)
\end{Preuve}

\noindent Implication from right to left follows from (\ref{equensures2a})\\
\begin{Preuve}
p\mapsto q\in \Ee(G)\equiv G\cdot x\in p\gg_{w} x\in q\\
\Com{\Imp}{ }\\
p\mapsto q\in \Ee(G)\Imp \exists G\cdot (G\in \S\And G\cdot x\in p\gg_{w} x\in q)\\
\Com{\Imp}{ }\\
\exists G\cdot (p\mapsto q\in \Ee(G))\Imp \exists G\cdot (G\in \S\And G\cdot x\in p\gg_{w} x\in q)\\
\Com{\Imp}{ }\\
p\mapsto q\in \Ew\Imp \exists G\cdot (G\in \S\And G\cdot x\in p\gg_{w} x\in q)
\end{Preuve}
\eop

\noindent{\bf Proof of (\ref{equensures1a})}
\begin{Preuve}
G\cdot P(x)\gg_{w} Q(x)\\
\Com{\equiv}{ def. $\gg_w$ }\\
\forall x\cdot (I(x)\And P(x)\And \neg Q(x)\Imp ([\sub(S)]\,(P(x)\Or Q(x)))\And \grd(\sub(G))\And [\sub(G)]\,Q(x))\\
\Com{\equiv}{ set theory }\\
\forall x\cdot (x\in \{\,z\,|\,I(z)\And P(z)\,\}\cap \{\,z\,|\,I(z)\And \neg Q(x)\,\}\Imp ([\sub(S)]\,(P(x)\Or Q(x)))\And \grd(\sub(G))\And [\sub(G)]\,Q(x))\\
\Com{\equiv}{ $I(x)\Imp [\sub(S)]\,I(x)$, conjunctive $S$ and $G$ }\\
\forall x\cdot (x\in \{\,z\,|\,I(z)\And P(z)\,\}\cap \{\,z\,|\,I(z)\And \neg Q(x)\,\}\Imp ([\sub(S)]\,(I(x)\And P(x)\Or I(x)\And Q(x)))\And \grd(\sub(G))\And [\sub(G)]\,I(x)\And Q(x))\\
\Com{\equiv}{ set transformers }\\
\forall x\cdot (x\in \{\,z\,|\,I(z)\And P(z)\,\}\cap \{\,z\,|\,I(z)\And \neg Q(x)\,\}\Imp x\in S(\{\,z\,|\,z\in u\And P(z)\,\}\cup \{\,z\,|\,z\in u\And Q(x)\,\})\And x\in \grs(G)\And x\in G(\{\,z\,|\,z\in u\And Q(x)\,\}))\\
\Com{\equiv}{ set theory, $u=\{\,z\,|\,I(z)\,\}$ }\\
\{\,z\,|\,z\in u\And P(z)\,\}\cap \cpl{\{\,z\,|\,z\in u\And Q(x)\,\}}\subseteq S(\{\,z\,|\,z\in u\And P(z)\,\}\cup \{\,z\,|\,z\in u\And Q(x)\,\})\cap \grs(G)\cap G(\{\,z\,|\,z\in u\And Q(x)\,\})\\
\Com{\equiv}{ def. $\Ee(G)$ }\\
\{\,z\,|\,z\in u\And P(z)\,\}\mapsto \{\,z\,|\,z\in u\And Q(x)\,\}\in \Ee(G)
\end{Preuve}
\eop

\noindent{\bf Proof of (\ref{equensures2a})}
\begin{Preuve}
p\mapsto q\in \Ee(G)\\
\Com{\equiv}{ def. $\Ew$ }\\
p\cap \cpl{q}\subseteq S(p\cup q)\cap \grs(G)\cap G(q)\\
\Com{\equiv}{ }\\
\forall x\cdot (x\in p\cap \cpl{q}\Imp x\in S(p\cup q)\cap \grs(G)\cap G(q))\\
\Com{\equiv}{ $x\in p\Imp I(x)$ }\\
\forall x\cdot (I(x)\And x\in p\And \neg x\in q\Imp x\in S(p\cup q)\cap \grs(G)\cap G(q))\\
\Com{\equiv}{ set transformers }\\
\forall x\cdot (I(x)\And x\in p\And \neg x\in q\Imp ([\sub(S)]\,(x\in p\Or x\in q))\And \grd(G)\And [\sub(G)]\,x\in q)\\
\Com{\equiv}{ def. $\gg_w$ }\\
G\cdot x\in p\gg_{w} x\in q
\end{Preuve}

\eop

\newpage

\newpage
\newcommand{\Fsb}{\F(\m{si}\cap b)}
\newcommand{\Wm}{W_{m}}
\section{Extension of Semantics to Consider the Strongest
    Invariant}\label{SI}

In this annex, the strongest invariant is considered in
definitions of \emph{termination\/} and \emph{reachability\/}
relations. It allows us to preserve soundness of {\sc unity} logic.
Certain definitions and proofs are independent of the strongest
invariant and they remain unchanged. New definitions and proofs are
only considered in this annex. It is structured in four parts.  In
section \ref{SecSI}, the strongest invariant is presented. In
section \ref{Sgeneral}, the general theorem about soundness and
completeness is restated and proved. In section \ref{Sminimal},
\emph{termination\/} and basic relation for \emph{leads-to\/} are
redefined to consider a minimal progress assumption, and the
hypothesis of the theorem of soundness and completeness are proved. In
section \ref{Sweak} an similar treatment is considered to the case of
a weak fairness assumption.

\subsection{Strongest Invariant}\label{SecSI}

Original definitions of the fundamental relations \emph{unless\/} and
\emph{ensures\/} in {\sc unity} logic \cite{ChMiPPD}, do not consider
initial conditions. On another hand, in order to give the possibility
to prove valid properties (completeness), in \cite{ChMiPPD} the
\emph{Substitution Axiom} is proposed. Basically, this axiom states
that any invariant predicate may be replaced by $\True$ and vice
versa.

Original definitions of the fundamental relations in \cite{ChMiPPD},
along with the substitution axiom, give an unsound proof system as it
is reported in \cite{SanESAU}. To fix this problem, the relation
\emph{unless\/} and \emph{ensures\/} are redefined to consider initial
conditions. The new definitions in \cite{SanESAU} consider the
\emph{strongest invariant}, which holds in the reachable states.
Moreover, the substitution axiom is replaced by a substitution rule,
which becomes a theorem in the new logic. However, in \cite{PraEUSR},
a problem with the new rule is reported and another substitution rule
is proposed. In this report we are not concerned by this last issue.

Following the proposal in \cite{SanESAU}, and adapting the definition
to our framework, the strongest invariant is defined as follows:

\begin{definition}
  \emph{Strongest Invariant}\\
  
  Let $\SS$ be a \B\ event system with state variable $x$,
  initialization $U$ and choice of events $S$.The strongest invariant
  \m{SI} of $\SS$ is the strongest predicate $X$ satisfying:
  \[ \forall x\cdot ((X \Imp [S]\,X) \And (([x':=x]\,\prd(U)) \Imp X))\]
\end{definition}
where $\prd(U)$ denotes the before-after relation associated with the
initilization $U$ \cite{AbrTBB}. Using the strongest invariant, the
definition of the \emph{ensures\/} relation, to specify basic liveness
properties, under our two fairness assumptions is as follows:

\begin{definition}
  \emph{Ensures under Minimal Progress}\\

  Let $\SS$ be a \B\ event system with state variable $x$,
  initialization $U$ and choice of events $S$. For any predicate $P$
  and $Q$ over $x$, the basic liveness relation $\gg_m$ is defined as
  follows: 
  \[ P \gg_{m} Q \equiv \m{SI} \And P \And \neg Q \Imp (([S]\; Q )
  \And \grd(S)) \]
\end{definition}

\newpage

\begin{definition}
  \emph{Ensures under Weak Fairness}\\
  
  Let $\SS$ be a \B\ event system with state variable $x$,
  initialization $U$, choice of events $S$ and $G$ an event of $\SS$.
  For any predicate $P$ and $Q$ over $x$, the basic liveness relation
  $\gg_w$ is defined as follows:
  \[ G\cdot P \gg_{w} Q \equiv \m{SI} \And P \And \neg Q \Imp (([S]\;
  P\Or Q ) \And ([G]\;Q) \And \grd(S)) \]
\end{definition}

Finally $\m{si}$, the set counterpart of the strongest invariant, is
given by the following definition
\[ \m{si} = \{ z \;|\; \m{SI}(z) \}\]

\subsection{General Framework}\label{Sgeneral}

The body of iteration in the general framework does not change:
\[\parbox{.937\textwidth}{$\F(r) = (\cpl{r}\Rarrow W)$\hfill(\ref{defF})}\]
The \emph{termination\/} relation is redefined to consider the
strongest invariant as follows:
\begin{definition}\label{defTS}
\emph{(Termination Relation)}
\begin{eqnarray}
\label{termS}
\lefteqn{\T=\{\,a\mapsto b\,|\,a\subseteq u\And b\subseteq u\And
  \m{si}\cap a\subseteq \fix(\F(\m{si}\cap b))\,\}} 
\end{eqnarray}
\end{definition}
The \emph{reachability\/} relation is not modified directly:
\begin{definition}\label{reach1}
  \emph{(Reachability Relation)}\\
  The \emph{reachability} relation $\L$, $\L\in \Pow(u)\Rel\Pow(u)$,
  is defined by the following
  induction scheme:\\[.5ex]
\noindent\emph{(\textbf{SBR})}: $\E\subseteq \L$\\[.25ex]
\emph{(\textbf{STR})}: $\L \!\sco \L \subseteq \L$\\[.25ex]
\emph{(\textbf{SDR})}: $\forall (q,l)\cdot (q\in\Pow(u)\And l\subseteq
\Pow(u) \Imp (l\times\{q\} \subseteq \L \Imp \bigcup(l)\mapsto
q \in \L))$ \\[.25ex]
\emph{\textbf{Closure}}: $\forall \m{l'}\cdot (l'\in u\Rel u \And
\E\subseteq \m{l'}\And \m{l'}\sco \m{l'}\subseteq \m{l'}\And$\\
$\forall (q,l)\cdot (q\in \Pow(u)\And l\subseteq \Pow(u)\And l\times
\{q\}\subseteq \m{l'}\Imp \bigcup(l)\mapsto q\in \m{l'})\Imp
\L\subseteq \m{l'})$
\end{definition}
However, when $\L$ is instantiated to minimal or weak fairness
assumptions, the strongest invariant is considered in the definition of
the base relation.

The theorem of soundness and completeness, taking into account  new
definitions of $\T$ and $\L$ remains basically unchanged. Only
hypothesis concerning the strongest invariant and the basic relation,
are modified with respect to the precedent version.

\begin{theorem}\label{STsoundcomp}
  \emph{(Soundness and Completeness)}\\
  Let $W$ be a monotonic and strict set transformer and $\F(r) =
  (\cpl{r}\Rarrow W)$ for any $r$ in $\Pow(u)$. Let relations $\T$ and
  $\L$ be defined as definitions \ref{defTS} and \ref{reach1}
  respectively. Considering (a) $a\mapsto b\in \E\Imp \m{si}\cap a\cap
  \cpl{b}\subseteq W(\m{si}\cap b)$, (b) $a\subseteq b \Imp a\mapsto b\in\E$, (c)
  $\m{si}\cap a\subseteq b\Imp a\mapsto b\in \E$ and (d) $W(r)\mapsto
  r\in\L$, for any $a$, $b$ and $r$ in $\Pow(u)$, the following
  equality holds:
  \[\L = \T\]
\end{theorem}
The proof is given in the following section.

In appendix \ref{equsetpred}, the equivalence between the reachability
relation $\L$ and the definition of the \emph{leads-to\/} relation
$\leadsto$ is proved. That proof does not consider the strongest
invariant in the given definitions. In order to connect the definition
of the reachability relation, considering the strongest invariant, a
demonstration similar to the proof of (\ref{defequlto}), can be given
to prove the following equivalence:
 \[
 \parbox{.937\textwidth}{$P(x)\leadsto Q(x)\equiv \{\,z\,|\,z\in
   \m{si}\And P(z)\,\}\mapsto \{\,z\,|\,z\in \m{si}\And Q(z)\,\}\in
   \L$\hfill}
\] 
which relates the definition of the \emph{leads-to\/} relation with
the reachability relation considering the strongest invariant.

\subsection{Proof of Soundness and Completeness}\label{Ssouncomp}
As before, the proof is divided in $\L\subseteq\T$ and
$\T\subseteq\L$.

\subsubsection{Proof of $\L\subseteq\T$}\mbox{}\\


The proof of $\L\subseteq\T$ follows from the closure clause in
definition of $\L$. According to this clause, $\T$ must contain the
base relation $\E$, and it must be transitive and disjunctive. The
following paragraphs present the proof of these cases.

\subsubsection*{Base Case}
\begin{eqnarray}
\label{gbc}
\lefteqn{a\mapsto b\in \E\Imp a\mapsto b\in \T}
\end{eqnarray}
\bop
\begin{PreuveL}
1 & a\mapsto b\in \E &  premise\\
2 & \m{si}\cap a\cap \cpl{b}\subseteq W(b\cap \m{si}) &  1 and hyp. (a)\\
3 & \m{si}\cap a\subseteq b\cup W(b\cap \m{si}) &  2\\
4 & \m{si}\cap a\subseteq \m{si}\cap b\cup W(b\cap \m{si}) &  3 \\
5 & \m{si}\cap a\subseteq \Fsb(b\cap \m{si}) &  4 and def. (\ref{defF})\\
6 & \m{si}\cap a\subseteq \Fsb^{2} &  5 and iterate (\ref{iterate})\\
7 & \m{si}\cap a\subseteq \fix(\Fsb) &  6 and (\ref{Pfix1})\\
8 & a\mapsto b\in \T &  7 and def. (\ref{termS})
\end{PreuveL}
\eop

\subsubsection*{Transitivity}
\begin{eqnarray}
\label{gtr}
\lefteqn{\forall (a,b)\cdot (a\mapsto b\in (\T\sco \T)\Imp a\mapsto b\in \T)}
\end{eqnarray}

\noindent The proof of (\ref{gtr}) requires the following property:
\begin{eqnarray}
\label{incfix1}
\lefteqn{\forall (a,b)\cdot (a\mapsto b\in \T\Imp \fix(\F(\m{si}\cap
  a))\subseteq \fix(\F(\m{si}\cap b)))} 
\end{eqnarray}
Taking $a\mapsto b\in\T$ as a premise, and considering
$\fix(\F(\m{si}\cap a))$ as the least fixpoint of $\F(\m{si}\cap a)$,
(\ref{incfix1}) follows from
$\F(\m{si}\cap a)(\fix(\F(\m{si}\cap b)))\subseteq \fix(\F(\m{si}\cap b))$ as follows:\\
\bop
\begin{PreuveL}
1 & \m{si}\cap a\subseteq \fix(\F(\m{si}\cap b)) &  from $a\mapsto b\in\T$\\
2 & \F(\m{si}\cap b)(\fix(\F(\m{si}\cap b)))=\fix(\F(\m{si}\cap b)) &  fixpoint definition\\
3 & W(\fix(\F(\m{si}\cap b)))\subseteq \fix(\F(\m{si}\cap b)) &  2 and (\ref{defF})\\
4 & \F(\m{si}\cap a)(\fix(\F(\m{si}\cap b)))\subseteq \fix(\F(\m{si}\cap b)) &  3, 1 and (\ref{defF})
\end{PreuveL}
\eop

\noindent{\bf Proof of (\ref{gtr})}
\begin{PreuveL}
1 & \exists c\cdot (a\mapsto c\in \T\And c\mapsto b\in \T) &  from $a\mapsto b\in\T ;\T$ \\
2 & \exists c\cdot (\m{si}\cap a\subseteq \fix(\F(\m{si}\cap c))\And c\mapsto b\in \T) &  3 and def. $\T$\\
3 & \exists c\cdot (\m{si}\cap a\subseteq \fix(\F(\m{si}\cap c))\And \fix(\F(\m{si}\cap c))\subseteq \fix(\F(\m{si}\cap b))) &  2 and (\ref{incfix1})\\
4 & \m{si}\cap a\subseteq \fix(\F(\m{si}\cap b)) &  3\\
5 & a\mapsto b\in \T &  6 and def. $\T$
\end{PreuveL}
\eop

\subsubsection*{Disjunction}
\begin{eqnarray}
\label{gdj}
\lefteqn{\forall (l,q)\cdot (l\times \{q\}\subseteq \T\Imp \bigcup(l)\mapsto q\in \T)}
\end{eqnarray}
\bop
\begin{PreuveL}
1 & l\times \{q\}\subseteq \T &  premise\\
2 & \forall p\cdot (p\in l\Imp p\mapsto q\in \T) &  1\\
3 & \forall p\cdot (p\in l\Imp \m{si}\cap p\subseteq \fix(\F(\m{si}\cap q))) &  2 and def. $\Tm$\\
4 & \bigcup p\cdot (p\in l\;|\;\m{si}\cap p)\subseteq \fix(\F(\m{si}\cap q)) &  3\\
5 & \m{si}\cap \bigcup p\cdot (p\in l\;|\;p)\subseteq \fix(\F(\m{si}\cap q)) &  4\\
6 & \m{si}\cap \bigcup(l)\subseteq \fix(\F(\m{si}\cap q)) &  5\\
7 & \bigcup(l)\mapsto q\in \T &  4 and def. $\Tm$
\end{PreuveL}
\eop

\subsubsection{Proof of $\T\subseteq\L$}\mbox{}\\
The proof of this inclusion requires the following property which is
already proved:
\[\parbox{.937\textwidth}{$\forall r\cdot (r\in \Pow(u)\Imp \F(r)^{\alpha}\mapsto r\in
  \L$\hfill(\ref{ordprop})}\]

\noindent{\bf Proof of $\T\subseteq\L$}
\begin{PreuveL}
1 & \m{si}\cap a\subseteq \fix(\Fsb) &  from $a\mapsto b\in\T$\\
2 & \exists \a\cdot (\m{si}\cap a\subseteq \Fsb^{\a}) &  1 and theorem 1\\
3 & \exists \a\cdot (a\mapsto \Fsb^{\a}\in \E) &  2 and hyp. (c)\\
4 & \exists \a\cdot (a\mapsto \Fsb^{\a}\in \L) &  3 and STR\\
5 & a\mapsto \m{si}\cap b\in \L &  4 and (\ref{ordprop})\\
6 & a\mapsto b\in \L &  5, $\m{si}\cap b\mapsto b\in \L$, STR
\end{PreuveL}
\eop
\subsection{Minimal Progress}\label{Sminimal}

The body of iteration under a minimal progress assumption does not
change:
\begin{eqnarray*}
\lefteqn{\Fm(r)=(\cpl{r}\Rarrow \Wm)}
\end{eqnarray*} 
where $\Wm$ remains defined as before:
\[\parbox{.937\textwidth}{$\Wm = \grs(S)\;|\;S$\hfill(\ref{Wmp})}\]
\emph{termination\/} relation under weak fairness assumption is
defined as follows:
\begin{eqnarray}
\label{termSmp}
\lefteqn{\Tm=\{\,a\mapsto b\,|\,a\subseteq u\And b\subseteq u\And
  \m{si}\cap a\subseteq \fix(\Fm(\m{si}\cap b))\,\}} 
\end{eqnarray}


Basic relation $\Em$ for \emph{reachability\/} relation $\Lm$ under minimal
progress assumptions is defined as follows:
\begin{eqnarray}
\label{defSEm}
\lefteqn{\Em=\{\,a\mapsto b\,|\,a\subseteq u\And b\subseteq u\And
  \m{si}\cap a\cap \cpl{b}\subseteq S(b)\cap \grs(S)\,\}} 
\end{eqnarray}
Relation $\Lm$ is defined by an induction scheme, according to
definition \ref{reach}.

The following proofs of hypothesis in theorem \ref{STsoundcomp} allow
us to conclude the equality between \emph{reachability\/} and
\emph{termination\/} relations under a minimal progress assumption:
\begin{eqnarray*}
\lefteqn{\Tm=\Lm}
\end{eqnarray*}

\subsubsection{Proof of Premise (a)}
\begin{eqnarray}
\label{mpa}
\lefteqn{a\mapsto b\in \Em\Imp \m{si}\cap a\cap \cpl{b}\subseteq \Wm(\m{si}\cap b)}
\end{eqnarray}
\bop
\begin{Preuve}
a\mapsto b\in \Em\\
\Com{\equiv}{ (\ref{defSEm}) }\\
\m{si}\cap a\cap \cpl{b}\subseteq S(b)\cap \grs(S)\\
\Com{\Imp}{ $\m{si}\subseteq S(\m{si})$, conjunctive $S$ }\\
\m{si}\cap a\cap \cpl{b}\subseteq S(\m{si}\cap b)\cap \grs(S)\\
\Com{\equiv}{ (\ref{Wmp}) }\\
\m{si}\cap a\cap \cpl{b}\subseteq \Wm(\m{si}\cap b)
\end{Preuve}
\eop
\subsubsection{Proof of Premise (b)}
\begin{eqnarray}
\label{mpb}
\lefteqn{a\subseteq b\Imp a\mapsto b\in \Em}
\end{eqnarray}
\bop
\begin{PreuveL}
1 & a\subseteq b &  premise\\
2 & \m{si}\cap a\cap \cpl{b}=\ets &  1\\
3 & \m{si}\cap a\cap \cpl{b}\subseteq \grs(S)\cap S(b) &  2\\
4 & a\mapsto b\in \Em &  3 and (\ref{defSEm})
\end{PreuveL}
\eop

\subsubsection{Proof of Premise (c)}
\begin{eqnarray}
\label{mpc}
\lefteqn{\m{si}\cap a\subseteq b\Imp a\mapsto b\in \Em}
\end{eqnarray}
\bop
\begin{PreuveL}
1 & \m{si}\cap a\subseteq b &  premise\\
2 & \m{si}\cap b\cap \cpl{b}=\ets &  1\\
3 & \m{si}\cap a\cap \cpl{b}\subseteq \grs(S)\cap S(b) &  2\\
4 & a\mapsto b\in \Em &  3 and (\ref{defSEm})
\end{PreuveL}
\eop

\subsubsection{Proof of Premise (d)}
\begin{eqnarray}
\label{mpd}
\lefteqn{\Wm(r)\mapsto r\in \Lm}
\end{eqnarray}
\bop
\begin{PreuveL}
1 & \m{si}\cap \grs(S)\cap S(r)\cap \cpl{r}\subseteq \grs(S)\cap S(r) &  trivial\\
2 & \grs(S)\cap S(r)\mapsto r\in \Em &  1 and def. $\Em$\\
3 & \Wm(r)\mapsto r\in \Em &  2 and (\ref{Wmp})\\
4 & \Wm(r)\mapsto r\in \Lm &  3 and def. $\Lm$
\end{PreuveL}
\eop

\subsection{Weak Fairness}\label{Sweak}
The body of iteration under weak fairness assumption does not change:
\[\parbox{.937\textwidth}{$\Fw(r) = \cpl{r}\Rarrow W_{w}$\hfill(\ref{defFW})}\]
where $\Ww$ is defined as follows:
\[\parbox{.937\textwidth}{$\Ww = \lambda r\cdot (r\subseteq u\;|\;\bigcup G\cdot
  (G\in\S\;|\;Y(r)(G)(r)))$\hfill(\ref{Wwf})}\]
and $Y(r)(G)(r)$ is defined by:
\[\parbox{.937\textwidth}{$Y(q)(G)(r)=\FIX(\cpl{q}\Rarrow (\grs(G)\cap G(r)\; |\; S))$\hfill(\ref{strY})}\]
\emph{termination\/} relation under weak fairness assumption is
defined as follows:
\begin{eqnarray}
\label{termSW}
\lefteqn{\Tw=\{\,a\mapsto b\,|\,a\subseteq u\And b\subseteq u\And
  \m{si}\cap a\subseteq \fix(\Fw(\m{si}\cap b))\,\}} 
\end{eqnarray}

Basic relation $\Ew$ for \emph{reachability\/} relation $\Lw$ under
weak fairness assumption is not changed directly:
\[\parbox{.937\textwidth}{$\Ew=\bigcup G\cdot (G\in \S\;|\;\Ee(G))$\hfill(\ref{defEw})}\]
However, $\Ee(G)$ is redefined to consider the strongest invariant:
\begin{eqnarray}
\label{defEew}
\lefteqn{\Ee(G)=\{\,a\mapsto b\,|\,a\subseteq u\And b\subseteq u\And
  \m{si}\cap a\cap \cpl{b}\subseteq S(a\cup b)\cap \cpl{G(\ets)}\cap
  G(b)\,\}} 
\end{eqnarray}
Relation $\Lw$ is defined by an induction scheme, according to
definition \ref{reach}.

The following proofs of hypothesis in theorem \ref{STsoundcomp} allow
us to conclude the equality between \emph{reachability\/} and
\emph{termination\/} relations under a weak fairness assumption:
\begin{eqnarray*}
\lefteqn{\Tw=\Lw}
\end{eqnarray*}

\subsubsection{Proof of Premise (a)}
\begin{eqnarray}
\label{wfa}
\lefteqn{a\mapsto b\in \Ew\Imp \m{si}\cap a\cap \cpl{b}\subseteq \Ww(\m{si}\cap b)}
\end{eqnarray}

The proof requires the following property:
\begin{eqnarray}
\label{propE'1}
\lefteqn{\forall G\cdot (G\in \S\And a\mapsto b\in \Ee(G)\Imp
  \m{si}\cap a\subseteq Y(\m{si}\cap b)(G)(\m{si}\cap b))} 
\end{eqnarray}
\bop
\begin{Preuve}
a\mapsto b\in \Ee(G)\\
\Com{\equiv}{ (\ref{defEew}) }\\
\m{si}\cap a\cap \cpl{b}\subseteq S(a\cup b)\cap \cpl{G(\ets)}\cap G(b)\\
\Com{\Imp}{ $\m{si}\subseteq S(\m{si})$, conjunctive $S$ and $G$ }\\
\m{si}\cap a\cap \cpl{b}\subseteq S(\m{si}\cap a\cup \m{si}\cap b)\cap \cpl{G(\ets)}\cap G(\m{si}\cap b)\\
\Com{\Imp}{ }\\
\m{si}\cap a\subseteq \m{si}\cap b\cup S(\m{si}\cap a\cup \m{si}\cap b)\cap \cpl{G(\ets)}\cap G(\m{si}\cap b)\\
\Com{\Imp}{ }\\
\m{si}\cap a\cup \m{si}\cap b\subseteq \m{si}\cap b\cup S(\m{si}\cap a\cup \m{si}\cap b)\cap \cpl{G(\ets)}\cap G(\m{si}\cap b)\\
\Com{\equiv}{ set transformers }\\
\m{si}\cap a\cup \m{si}\cap b\subseteq (\cpl{\m{si}\cap b}\Rarrow \grs(G)\cap G(\m{si}\cap b)\; |\; S)(\m{si}\cap a\cup \m{si}\cap b)\\
\Com{\Imp}{ greatest fixpoint property }\\
\m{si}\cap a\subseteq \FIX(\cpl{\m{si}\cap b}\Rarrow \grs(G)\cap G(\m{si}\cap b)\; |\; S)\\
\Com{\equiv}{ (\ref{strY}) }\\
\m{si}\cap a\subseteq Y(\m{si}\cap b)(G)(\m{si}\cap b)
\end{Preuve}
\eop

\noindent{\bf Proof of (\ref{wfa})}
\begin{PreuveL}
1 & \forall G\cdot (G\in \S\Imp (a\mapsto b\in \Ee(G)\Imp \m{si}\cap a\subseteq Y(\m{si}\cap b)(G)(\m{si}\cap b))) &  (\ref{propE'1})\\
2 & \exists G\cdot (G\in \S\And a\mapsto b\in \Ee(G))\Imp \m{si}\cap a\subseteq W_{w}(\m{si}\cap b) &  1 and (\ref{Wwf}). \\
3 & a\mapsto b\in \bigcup G\cdot (G\in \S\;|\;\Ee(G))\Imp \m{si}\cap a\subseteq W_{w}(\m{si}\cap b) &  2\\
4 & a\mapsto b\in \Ew\Imp \m{si}\cap a\subseteq W_{w}(\m{si}\cap b) &  3 and (\ref{defEw})\\
5 & a\mapsto b\in \Ew\Imp \m{si}\cap a\subseteq \m{si}\cap b\cup
W_{w}(\m{si}\cap b) &  4, $\m{si}\cap b\!\subseteq\!W_{w}(\m{si}\cap b)$ \\
6 & a\mapsto b\in \Ew\Imp \m{si}\cap a\cap \cpl{b}\subseteq W_{w}(\m{si}\cap b) &  5
\end{PreuveL}
\eop

\subsubsection{Proof of Premise (b)}
\begin{eqnarray}
\label{wfb}
\lefteqn{a\subseteq b\Imp a\mapsto b\in \Ew}
\end{eqnarray}
\bop
\begin{PreuveL}
1 & a\subseteq b &  premise\\
2 & \m{si}\cap a\cap \cpl{b}=\ets &  1\\
3 & \forall G\cdot (G\in \S\Imp \m{si}\cap a\cap \cpl{b}\subseteq S(a\cup b)\cap \cpl{G(\ets)}\cap G(b)) &  2\\
4 & \forall G\cdot (G\in \S\Imp a\mapsto b\in \Ee(G)) &  3 and (\ref{defEew})\\
5 & a\mapsto b\in \bigcup G\cdot (G\in \S\;|\;\Ee(G)) &  4\\
6 & a\mapsto b\in \Ew &  5 (\ref{defEw})
\end{PreuveL}
\eop

\subsubsection{Proof of Premise (c)}
\begin{eqnarray}
\label{wfc}
\lefteqn{\m{si}\cap a\subseteq b\Imp a\mapsto b\in \Ew}
\end{eqnarray}
\bop
\begin{PreuveL}
1 & \m{si}\cap a\subseteq b &  premise\\
2 & \m{si}\cap a\cap \cpl{b}=\ets &  1\\
3 & \forall G\cdot (G\in \S\Imp \m{si}\cap a\cap \cpl{b}\subseteq S(a\cup b)\cap \cpl{G(\ets)}\cap G(b)) &  2\\
4 & \forall G\cdot (G\in \S\Imp a\mapsto b\in \Ee(G)) &  3 and (\ref{defEew})\\
5 & a\mapsto b\in \bigcup G\cdot (G\in \S\;|\;\Ee(G)) &  4\\
6 & a\mapsto b\in \Ew &  5 (\ref{defEw})
\end{PreuveL}
\eop

\subsubsection{Proof of Premise (d)}
\begin{eqnarray}
\label{wfd}
\lefteqn{\forall r\cdot (r\subseteq u\Imp \Ww(r)\mapsto r\in \Lw)}
\end{eqnarray}

The proof requires the following property:
\begin{eqnarray}
\label{propEwp1}
\lefteqn{\forall (G,r)\cdot (G\in \S\And r\subseteq u\Imp
  Y(r)(G)(r)\mapsto r\in \Ee(G))} 
\end{eqnarray}
\bop
\begin{PreuveL}
1 & S(Y(r)(G)(r))\subseteq S(Y(r)(G)(r)\cup r) & monotony of $S$ \\
2 & Y(r)(G)(r)=r\cup \grs(G)\cap G(r)\cap S(Y(r)(G)(r)) &  (\ref{strY})\\
3 & Y(r)(G)(r)\cap \cpl{r}\subseteq \grs(G)\cap G(r)\cap S(Y(r)(G)(r)\cup r) & 2, 1\\
4 & \m{si}\cap Y(r)(G)(r)\cap \cpl{r}\subseteq Y(r)(G)(r)\cap \cpl{r} &  trivial\\
5 & Y(r)(G)(r)\mapsto r\in \Ee(G) &  4, 3 and (\ref{defEw})
\end{PreuveL}
\eop

\noindent{\bf Proof of (\ref{wfd})}
\begin{PreuveL}
1 & \forall G\cdot (G\in \S\Imp Y(r)(G)(r)\mapsto r\in \Ee(G)) &  from (\ref{propEwp1}) \\
2 & \forall G\cdot (G\in \S\Imp \Ee(G)\subseteq \Ew) &  def. $\Ew$\\
3 & \forall G\cdot (G\in \S\Imp Y(r)(G)(r)\mapsto r\in \Ew) &  2 and 1\\
4 & \forall G\cdot (G\in \S\Imp Y(r)(G)(r)\mapsto r\in \Lw) &  def. $\Lw$\\
5 & \{\,Y(r)(G)(r)\,|\,G\in \S\,\}\times \{r\}\subseteq \Lw &  4\\
6 & \bigcup(\{\,Y(r)(G)(r)\,|\,G\in \S\,\})\mapsto r\in \Lw &  5 and SDR\\
7 & \bigcup G\cdot (G\in \S\;|\;Y(r)(G)(r))\mapsto r\in \Lw &  6\\
8 & \Ww(r)\mapsto r\in \Lw &  7 and def. $F$
\end{PreuveL}
\eop




\newpage
\section{Proofs of section \ref{souncomp}}\label{AppendSounComp}
\subsection[Proof of (\ref{Pfix1})]{Proof of (\ref{Pfix1}): $\forall \a\cdot  (f^{\a}\subseteq \fix(f))$}

\bop
\noindent Successor ordinal:
\begin{Preuve}
f^{\a}\subseteq \fix(f)\\
\Com{\Imp}{ }\\
f(f^{\a})\subseteq f(\fix(f))\\
\Com{\Imp}{ }\\
f^{\a+1}\subseteq \fix(f)
\end{Preuve}
\noindent Limit ordinal:
\begin{Preuve}
\forall \b\cdot (\b< \a\Imp f^{\b}\subseteq \fix(f))\\
\Com{\Imp}{ monotony }\\
\forall \b\cdot (\b< \a\Imp f(f^{\b})\subseteq f(\fix(f)))\\
\Com{\Imp}{ fixpoint def.}\\
\forall \b\cdot (\b< \a\Imp f(f^{\b})\subseteq \fix(f))\\
\Com{\Imp}{ }\\
\bigcup \b\cdot (\b< \a\;|\;f(f^{\b}))\subseteq \fix(f)\\
\Com{\equiv}{ def. iterate }\\
f^{\a}\subseteq \fix(f)
\end{Preuve}
\eop

\subsection[Proof of (\ref{ordprop})]{Proof of (\ref{ordprop}): $\F(r)^{\a}\mapsto r\in \L\Imp \F(r)^{\a+1}\mapsto r\in \L$}

\bop
\noindent Successor ordinal
\begin{PreuveL}
1 & \F(r)^{\a}\mapsto r\in L &  ind. hyp.\\
2 & W(\F(r)^{\a})\mapsto \F(r)^{\a}\in L & from hyp. (c) th. \ref{Tsoundcomp}\\
3 & r\mapsto r\in L &hyp. (b) th. \ref{Tsoundcomp}, SBR\\
4 & r\cup W(\F(r)^{\a})\mapsto r\in L &  3, 2 and SDR\\
6 & \F(r)(\F(r)^{\a})\mapsto r\in L &  5 and (\ref{defF})\\
7 & \F(r)^{\a+1}\mapsto r\in L &  6 and def. iterate 
\end{PreuveL}
\eop

\section{Proofs of Section \ref{weak}}\label{AppendWeak}

\subsection[Termination Set of Fair Loop]{Termination Set of Fair Loop: $\pre(Y(q)(G)) =
  \fix(\cpl{q}\cap G(\ets)\Rarrow (\cpl{S(q)}\; |\; S))$}

\bop
\begin{Preuve}
\pre(Y(q)(G))\\
\Com{=}{ def. of $\pre$ }\\
Y(q)(G)(u)\\
\Com{=}{ (\ref{defY}) }\\
q\cup ((S\sco Y(q)(G))\dtl (\grs(G)\; |\; G))(u)\\
\Com{=}{ (\ref{predtl}), $X=(S\sco Y(q)(G))$, $Z=(\grs(G)\; |\; G)$ }\\
q\cup ((X(u)\cup Z(u))\cap (\cpl{X(\ets)}\cup Z(u))\cap (X(u)\cup \cpl{Z(\ets)}))\\
\Com{=}{ $G(u)=u$, $Z(u)=\grs(G)$, $\cpl{Z(\ets)}=u$ }\\
q\cup ((X(u)\cup \grs(G))\cap (\cpl{X(\ets)}\cup \grs(G)))\\
\Com{=}{ distributivity }\\
q\cup \grs(G)\cup (X(u)\cap \cpl{X(\ets)})\\
\Com{=}{ $X=(S\sco Y(q)(G))$, set transformer }\\
q\cup \grs(G)\cup (S(Y(q)(G)(u))\cap \cpl{S(Y(q)(G)(\ets))})\\
\Com{=}{ def. $\grs(Y(q)(G))$, $\grs(G)$ and set transformer }\\
(\cpl{q}\cap G(\ets)\Rarrow (\cpl{S(q)}\; |\; S))(Y(q)(G)(u))\\
\Com{=}{ extreme solution of recursive equation }\\
\fix(\cpl{q}\cap G(\ets)\Rarrow (\cpl{S(q)}\; |\; S))
\end{Preuve}
\eop

\subsection[Liberal Precondition of Fair Loop]{Liberal of $Y(q)(G)$: $\Li(Y(q)(G))(r)=\FIX(\cpl{q}\Rarrow (\grs(G)\cap G(r)\;|\;S))$}

\noindent For $r\subseteq u\And r\neq u$: 
\begin{Preuve}
\Li(Y(q)(G))(r)\\
\Com{=}{ (\ref{defY}), Liberal set transformer of guard and dovetail }\\
q\cup \Li(S\sco Y(q))(r)\cap \Li(\grs(G)\; |\; G)(r)\\
\Com{=}{ $r\neq u$, $\Li(\grs(G)\; |\; G)(r)=\grs(G)\cap \Li(G)(r)$,$\Li(G)(r)=G(r)$ }\\
q\cup \Li(S\sco Y(q))(r)\cap \grs(G)\cap G(r)\\
\Com{=}{ Liberal set transformer of sequencing, $\Li(S)(r)=S(r)$ }\\
q\cup S(\Li(Y(q))(r))\cap \grs(G)\cap G(r)\\
\Com{=}{ Liberal set transformer of guarded and preconditioned events }\\
(\cpl{q}\Rarrow \grs(G)\cap G(r)\; |\; S)(\Li(Y(q)(G))(r))\\
\Com{=}{ extreme solution of recursive equation }\\
\FIX(\cpl{q}\Rarrow \grs(G)\cap G(r)\; |\; S)
\end{Preuve}

\noindent For $r=u$ we prove:
\[ \Li(X(q)(G))(u)=u\]
\noindent First, we note that equality $\Li(X(q)(G))(u)=\FIX(\cpl{q}\Rarrow S)$
holds:
\begin{Preuve}
\Li(Y(q)(G))(u)\\
\Com{=}{ (\ref{defY}), Liberal set transformer of guard and dovetail }\\
q\cup \Li(S\sco Y(q))(u)\cap \Li(\grs(G)\; |\; G)(u)\\
\Com{=}{ $\Li(\grs(G)\; |\; G)(u)=u$ }\\
q\cup \Li(S\sco Y(q))(u)\\
\Com{=}{ $S(u)=u$, set transformers }\\
(\cpl{q}\Rarrow S)(\Li(Y(q)(G))(u))\\
\Com{=}{ extreme solution }\\
\FIX(\cpl{q}\Rarrow S)
\end{Preuve}
As $(\cpl{q}\Rarrow S)(u)=u$  holds, it follows: $u\subseteq
\FIX(\cpl{q}\Rarrow S)$. Therefore, $\Li(X(q)(G))(u)=u$ follows from
$u\subseteq\FIX(\cpl{q}\Rarrow S)$ and equality.
\eop

\subsection[Proof of (\ref{libInpre})]{Proof of (\ref{libInpre}): $\Li(Y(q)(G))(r)\subseteq \pre(Y(q)(G))$}

\begin{Preuve}
\Li(Y(q)(G))(r)\\
\Com{=}{ (\ref{LiY}) and fixpoint property }\\
q\cup \grs(G)\cap G(r)\cap S(\Li(Y(q)(G))(r))\\
\Com{\subseteq}{ set theory }\\
q\cup \grs(G)\\
\Com{\subseteq}{ set theory }\\
q\cup \grs(G)\cup \cpl{S(q)}\cap S(\pre(Y(q)(G)))\\
\Com{=}{set transformers }\\
(\cpl{q}\cap G(\ets)\Rarrow (\cpl{S(q)}\; |\; S))(\pre(Y(q)(G)))\\
\Com{=}{ (\ref{preY}) and fixpoint property }\\
\pre(Y(q)(G))
\end{Preuve}
\eop

\subsection[Monotonicity of Fair Loop]{Monotonicity of Fair Loop: $a\subseteq b\Imp Y(q)(G)(a)\subseteq Y(q)(G)(b))$} 

\noindent Let $a$ and $b$ be two subsets of $u$, $F(q)(a)$ and $F(q)(b)$ be the following set transformers:
\begin{eqnarray*}
\lefteqn{F(q)(a)=(\cpl{q}\Rarrow G(a)\cap \grs(G)\; |\; S)}\\
\lefteqn{F(q)(b)=(\cpl{q}\Rarrow G(b)\cap \grs(G)\; |\; S)}
\end{eqnarray*}

\bop
\begin{Preuve}
a\subseteq b\\
\Com{\Imp}{ Monotonicity of $G$ }\\
G(a)\subseteq G(b)\\
\Com{\Imp}{ set theory }\\
\forall r\cdot (r\subseteq u\Imp (q\cup G(a)\cap \grs(G)\cap S(r))\subseteq (q\cup G(b)\cap \grs(G)\cap S(r)))\\
\Com{\equiv}{ set transformers }\\
\forall r\cdot (r\subseteq u\Imp (\cpl{q}\Rarrow G(a)\cap \grs(G)\; |\; S)(r)\subseteq (\cpl{q}\Rarrow G(b)\cap \grs(G)\; |\; S)(r))\\
\Com{\equiv}{ def. $F(q)(a)$ and $F(q)(b)$ }\\
\forall r\cdot (r\subseteq u\Imp F(q)(a)(r)\subseteq F(q)(b)(r))\\
\Com{\Imp}{ set theory }\\
\forall r\cdot (r\subseteq u\Imp (r\subseteq F(q)(a)(r))\Imp r\subseteq F(q)(b)(r))\\
\Com{\Imp}{ set theory }\\
\{\,r\,|\,r\subseteq u\And r\subseteq F(q)(a)(r)\,\}\subseteq \{\,r\,|\,r\subseteq u\And r\subseteq F(q)(b)(r)\,\}\\
\Com{\Imp}{ set theory }\\
\bigcup(\{\,r\,|\,r\subseteq u\And r\subseteq F(q)(a)(r)\,\})\subseteq \bigcup(\{\,r\,|\,r\subseteq u\And r\subseteq F(q)(b)(r)\,\})\\
\Com{\Imp}{ def. $\FIX(f)$, $F(q)(a)$ and $F(q)(b)$ }\\
\FIX(\cpl{q}\Rarrow G(a)\cap \grs(G)\; |\; S)\subseteq \FIX(\cpl{q}\Rarrow G(b)\cap \grs(G)\; |\; S)\\
\Com{\equiv}{ (\ref{strY}) }\\
Y(q)(G)(a)\subseteq Y(q)(G)(b)
\end{Preuve}
\eop

\subsection[Guard of Fair Loop]{Guard of Fair Loop: $\grs(Y(q)(G)) = \cpl{q}$}

\begin{Preuve}
\grs(Y(q)(G))\\
\Com{=}{ def. guard }\\
\cpl{Y(q)(G)(\ets)}\\
\Com{=}{ (\ref{defY}) }\\
\cpl{q\cup ((S\sco Y(q)(G))\dtl (\grs(G)\; |\; G))(\ets)}\\
\Com{=}{ set theory }\\
\cpl{q}\cap \cpl{((S\sco Y(q)(G))\dtl (\grs(G)\; |\; G))(\ets)}\\
\Com{=}{ def. guard dovetail }\\
\cpl{q}\cap (\grs(S\sco Y(q)(G))\cup \grs(\grs(G)\; |\; G))\\
\Com{=}{ $\grs(\grs(G)\; |\; G)=u$ }\\
\cpl{q}
\end{Preuve}
\eop

\vspace{-.5cm}
\subsection[Strictness of $W_{w}$]{Strictness of $W_{w}$: $W_{w}(\ets) = \ets$}
\begin{minipage}{\textwidth}
\begin{Preuve}
W_{w}(\ets)\\
\Com{=}{ (\ref{Wwf}) }\\
\bigcup G\cdot (G\in \S\;|\;Y(\ets)(G)(\ets))\\
\Com{=}{ def. of $\grs$ }\\
\bigcup G\cdot (G\in \S\;|\;\cpl{\grs(Y(\ets)(G))})\\
\Com{=}{ $\grs(Y(\ets)(G))=\cpl{\ets}$ }\\
\bigcup G\cdot (G\in \S\;|\;\cpl{\cpl{\ets}})\\
\Com{=}{ }\\
\ets
\end{Preuve}
\eop
\end{minipage}

\subsection[Monotonicity of $W_{w}$]{Monotonicity of $W_{w}$: $a\subseteq b\Imp W_{w}(a)\subseteq W_{w}(b)$}

\noindent First we prove, for any subset $a$ and $b$ of $u$, $a\subseteq b\Imp
Y(a)(G)(b)\subseteq Y(b)(G)(b)$. 

\noindent Let $T(a) = \FIX(\cpl{a}\Rarrow
(\grs(G)\And G(b)\;|\;S))$:
\begin{Preuve}
a\subseteq b\\
\Com{\Imp}{ for any $G\in\S$  }\\
a\cup \grs(G)\cap G(b)\cap S(T(a))\subseteq b\cup \grs(G)\cap G(b)\cap
S(T(a))\\ 
\Com{\equiv}{ prop. $\FIX$ }\\
T(a)\subseteq b\cup \grs(G)\cap G(b)\cap S(T(a))\\
\Com{\equiv}{ guarded set transformer }\\
T(a)\subseteq (\cpl{b}\Rarrow \grs(G)\cap G(b)\; |\; S)(T(a))\\
\Com{\Imp}{ prop. $\FIX$ }\\
T(a)\subseteq \FIX(\cpl{b}\Rarrow \grs(G)\cap G(b)\; |\; S)\\
\Com{\equiv}{ (\ref{strY}) }\\
Y(a)(G)(b)\subseteq Y(b)(G)(b)
\end{Preuve}

\noindent Now, the proof of monotonicity is:
\begin{Preuve}
a\subseteq b\\
\Com{\Imp}{ monotonicity of $Y(q)(G)$ for $q=a$ and  $G\in\S$ }\\
Y(a)(G)(a)\subseteq Y(a)(G)(b)\\
\Com{\Imp}{ $Y(a)(G)(b)\subseteq Y(b)(G)(b)$ }\\
Y(a)(G)(a)\subseteq Y(b)(G)(b)\\
\Com{\Imp}{ }\\
Y(a)(G)(a)\subseteq \bigcup \m{G'}\cdot (\m{G'}\in \S\;|\;Y(b)(\m{G'})(b))\\
\Com{\Imp}{ }\\
\bigcup G\cdot (G\in \S\;|\;Y(a)(G)(a))\subseteq \bigcup \m{G'}\cdot (\m{G'}\in \S\;|\;Y(b)(\m{G'})(b))\\
\Com{\equiv}{ (\ref{Wwf}) }\\
W_{w}(a)\subseteq W_{w}(b)
\end{Preuve}
\eop

\newpage
\section{Proofs of section \ref{deriving}}\label{AppendDeriving}

\subsection[Proof of (\ref{varthA})]{Proof of (\ref{varthA}): $\forall n\cdot (n\in \Nat\Imp \m{v'}(n)=\bigcup i\cdot (i\in \Nat\And i< n\;|\;v(i)))$}


\noindent The proof is by induction over $\Nat$. The base case:
\begin{Preuve}
\m{v'}(0)\\
\Com{=}{ def. $v'$ }\\
\{\,z\,|\,z\in u\And V(z)< 0\,\}\\
\Com{=}{ $V\in u\Totfunc \Nat$ }\\
\ets\\
\Com{=}{ empty range }\\
\bigcup i\cdot (i\in \Nat\And i< 0\;|\;v(i))
\end{Preuve}

\noindent Inductive step:
\begin{Preuve}
\m{v'}(n)=\bigcup i\cdot (i\in \Nat\And i< n\;|\;v(i))\\
\Com{\Imp}{ }\\
\m{v'}(n)\cup v(n)=\bigcup i\cdot (i\in \Nat\And i< n\;|\;v(i))\cup v(n)\\
\Com{\Imp}{ def. $v$ and $v'$ }\\
\m{v'}(n+1)=\bigcup i\cdot (i\in \Nat\And i< n+1\;|\;v(i))
\end{Preuve}
\eop

\subsection[Proof of (\ref{varthB})]{Proof of (\ref{varthB}): $\bigcup i\cdot (i\in \Nat\;|\;\m{v'}(i+1))=\bigcup i\cdot (i\in \Nat\;|\;v(i))$}

\begin{Preuve}
\bigcup i\cdot (i\in \Nat\;|\;\m{v'}(i+1))\\
\Com{=}{ def. $v'$ }\\
\bigcup i\cdot (i\in \Nat\;|\;\{\,z\,|\,z\in u\And V(z)< i+1\,\})\\
\Com{=}{  }\\
\bigcup i\cdot (i\in \Nat\;|\;\{\,z\,|\,z\in u\And V(z)\leq i\,\})\\
\Com{=}{  }\\
\bigcup i\cdot (i\in \Nat\;|\;\{\,z\,|\,z\in u\And V(z)=i\,\})\\
\Com{=}{ def. $v$ }\\
\bigcup i\cdot (i\in \Nat\;|\;v(i))
\end{Preuve}
\eop

\subsection[Proof of (\ref{varthC})]{Proof of (\ref{varthC}): $\bigcup i\cdot (i\in \Nat\;|\;v(i))=u$}

\begin{Preuve}
u\subseteq \bigcup i\cdot (i\in \Nat\;|\;v(i))\\
\Com{\Follows}{ set. theory }\\
\forall x\cdot (x\in u\Imp x\in \bigcup i\cdot (i\in \Nat\;|\;v(i)))\\
\Com{\equiv}{ }\\
\forall x\cdot (x\in u\Imp \exists i\cdot (i\in \Nat\And x\in v(i)))\\
\Com{\equiv}{ def. $v$ }\\
\forall x\cdot (x\in u\Imp \exists i\cdot (i\in \Nat\And V(x)=i))\\
\Com{\Follows}{ }\\
V\in u\Totfunc \Nat
\end{Preuve}
\eop

\subsection[Proof of (\ref{varth2})]{Proof of (\ref{varth2}): $\forall n\cdot (n\in \Nat\Imp \bigcup i\cdot (i\in \Nat\And i\leq n\;|\;v(i))\cap p\subseteq f^{n+1})$}

\noindent The proof is by induction. Base case:
\begin{PreuveL}
1 & \forall n\cdot (n\in \Nat\Imp v(n)\cap p\subseteq f(\m{v'}(n))) &  premise\\
2 & v(0)\cap p\subseteq f(\m{v'}(0)) &  1\\
3 & v(0)\cap p\subseteq f(\ets) &  2 and def $v'$\\
4 & v(0)\cap p\subseteq f^{1} &  $f(\ets)=f^{1}$ (\ref{iterate})\\
5 & \bigcup i\cdot (i\in \Nat\And i\leq 0\;|\;v(i))\cap p\subseteq f^{0+1} &  4
\end{PreuveL}

\noindent Inductive step:
\begin{PreuveL}
1 & \bigcup i\cdot (i\in \Nat\And i\leq n\;|\;v(i))\cap p\subseteq f^{n+1} &  Inductive hyp.\\
2 & f(\bigcup i\cdot (i\in \Nat\And i\leq n\;|\;v(i))\cap p)\subseteq
f(f^{n+1}) &  1 and monotonic $f$\\
3 & \forall n\cdot (n\in \Nat\Imp v(n)\cap p\subseteq f(\m{v'}(n))) &  premise\\
4 & v(n+1)\cap p\subseteq f(\m{v'}(n+1)) &  3\\
5 & \m{v'}(n+1)=\bigcup i\cdot (i\in \Nat\And i\leq n\;|\;v(i)) &  from (\ref{varthA})\\
6 & p\subseteq f(p) &  premise\\
7 & v(n+1)\cap p\subseteq f(\m{v'}(n+1))\cap f(p) &  6 and 4\\
8 & v(n+1)\cap p\subseteq f(\m{v'}(n+1)\cap p) &  7 and conjunct. $f$\\
9 & v(n+1)\cap p\subseteq f(f^{n+1}) &  8, 5 and 2\\
10 & f^{n+1}\subseteq f^{n+2} &  from (\ref{iterate})\\
11 & \bigcup i\cdot (i\in \Nat\And i\leq n\;|\;v(i))\cap p\subseteq f^{n+2} &  10 and 1\\
12 & \bigcup i\cdot (i\in \Nat\And i\leq n\;|\;v(i))\cap p\cup v(n+1)\cap p\subseteq f^{n+2} &  11, 9 and (\ref{iterate})\\
13 & \bigcup i\cdot (i\in \Nat\And i\leq n+1\;|\;v(i))\cap p\subseteq f^{n+2} &  12
\end{PreuveL}
\eop

\subsection[Proof of (\ref{wfmp1})]{Proof of (\ref{wfmp1}): $\forall
  \a\cdot (\Fw(b)^{\a}\subseteq b\cup (\grs(S)\cap S(\fix(\Fw(b))))$}
The proof is given by transfinite induction considering the following
abbreviations:
\begin{eqnarray}
\label{defB}
\lefteqn{B=\fix(\Fw(b))}\\
\label{defFF}
\lefteqn{F=\Fw(b)}
\end{eqnarray}
Successor ordinal:
\begin{PreuveCL}{N}
1 & \Fwb^{\a}\subseteq b\cup (\grs(S)\cap S(B)) &;  Ind. Hyp.\\
2 & \Fwb(\Fi)=b\cup \Ww(\Fi) &;  (\ref{defFW})\\
3 & \Fwb(\Fi)=b\cup \bigcup G\cdot (G\in \S\;|\;Y(\Fi)(G)(\Fi)) &;  2 and (\ref{Wwf})\\
4 & \mbox{$\Fwb(\Fi)=b\cup \bigcup G\cdot (G\in \S\;|\;\FIX(\cpl{\Fi}\Rarrow \grs(G)\cap G(\Fi)\; |\; S))$} &\Esp{.5cm};  3 and (\ref{strY})\\
5 & \mbox{$\Fwb(\Fi)=b\cup \bigcup G\cdot (G\in \S\;|\;\Fi\cup (\grs(G)\cap G(\Fi)\cap S(Y(\Fi)(G)(\Fi))))$} &\Esp{2.5cm};  4\\
6 & \mbox{$\Fwb(\Fi)\subseteq b\cup \bigcup G\cdot (G\in \S\;|\;\Fi\cup (\grs(G)\cap S(Y(\Fi)(G)(\Fi)))) $}&\Esp{1cm}; 5\\
7 & \Fwb(\Fi)\subseteq b\cup \bigcup G\cdot (G\in \S\;|\;\Fi\cup (\grs(G)\cap S(\Ww(\Fi)))) &; 6 and (\ref{Wwf})\\
8 & \Fwb(\Fi)\subseteq b\cup \bigcup G\cdot (G\in \S\;|\;\Fi\cup (\grs(S)\cap S(\Ww(\Fi)))) &; 7 $\grs(G)\subseteq \grs(S)$ \\
9 & \Fwb(\Fi)\subseteq b\cup \bigcup G\cdot (G\in \S\;|\;\Fi\cup (\grs(S)\cap S(\Fwb(\Fi)))) &; 8 and (\ref{defFW})\\
10 & \Fwb^{\a+1}\subseteq b\cup \Fi\cup (\grs(S)\cap S(\Fwb^{\a+1})) &; 9 and (\ref{iterate})\\
11 & \Fwb^{\a+1}\subseteq b\cup \Fi\cup (\grs(S)\cap S(B)) &; 10 and (\ref{iterate})\\
12 & \Fwb^{\a+1}\subseteq b\cup (\grs(S)\cap S(B)) &; 11 and 1
\end{PreuveCL}
Limit ordinal:
\begin{PreuveCL}{N}
1 & \forall \b\cdot (\b< \a\Imp \FF^{\b}\subseteq b\cup \grs(S)\cap S(B)) &;  Ind. Hyp\\
2 & \FF(\FF^{\b})=b\cup \Ww(\FF^{\b}) &;  (\ref{defFW}), $\b<\a$\\
3 & \FF(\FF^{\b})=b\cup \bigcup G\cdot (G\in \S\;|\;Y(\FF^{\b})(G)(\FF^{\b})) &;  (\ref{Wwf}) \\
4 & \FF(\FF^{\b})=b\cup \bigcup G\cdot (G\in \S\;|\;\FIX(\cpl{\FF^{\b}}\Rarrow \grs(G)\cap G(\FF^{\b})\; |\; S)) &;  3, (\ref{strY})\\
5 & \mbox{$\FF(\FF^{\b})=b\cup \bigcup G\cdot (G\in \S\;|\;\FF^{\b}\cup (\grs(G)\cap G(\FF^{\b})\cap S(Y(\FF^{\b})(G)(\FF^{\b}))))$} &\Esp{2cm}; 4\\
6 & \FF(\FF^{\b})\subseteq b\cup \FF^{\b}\cup (\grs(S)\cap S(\FF(\FF^{\b}))) &;  5, (\ref{Wwf}) and (\ref{defFW})\\
7 & \FF(\FF^{\b})\subseteq b\cup \FF^{\b}\cup (\grs(S)\cap S(B)) &;  6 and (\ref{Pfix1})\\
8 & \FF(\FF^{\b})\subseteq b\cup (\grs(S)\cap S(B)) &;  7, 1 $\b<\a$\\
9 & \bigcup \b\cdot (b< \a\;|\;\FF(\FF^{\b}))\subseteq b\cup (\grs(S)\cap S(B)) &;  8\\
10 & \FF^{\a}\subseteq b\cup (\grs(S)\cap S(B)) &;  9 and (\ref{iterate})
\end{PreuveCL}

\renewcommand{\Fwb}{\Fw(b)}
Using (\ref{wfmp1}), the proof of sufficient conditions is as
follows:\\
\bop
\begin{PreuveL}
1 & \forall n\cdot (n\in \Nat\Imp \cpl{b}\cap v(n)\subseteq S(\m{v'}(n))) &  premise\\
2 & a\mapsto b\in \Lw &  premise\\
3 & \forall \a'\cdot (\Fwb^{\a'}\subseteq b\cup (\grs(S)\cap S(\fix(\Fwb)))) &  (\ref{wfmp1})\\
4 & \exists \a\cdot (\fix(\Fw(b))=\Fwb^{\a}) &  Theorem \ref{tglim}\\
5 & \fix(\Fwb)\subseteq b\cup (\grs(S)\cap S(\fix(\Fwb))) &  4 and 3\\
6 & \fix(\Fwb)\subseteq \Fm(b)(\fix(\Fwb)) &  5 and (\ref{defFm}) \\
7 & a\mapsto b\in \Tw &  2, equality (\ref{equalWF})   \\
8 & a\subseteq \fix(\Fwb) &  7 and def. $\Tw$\\
9 & \fix(\Fwb)\cap \cpl{b}\subseteq \grs(S) &  5\\
10 & \forall n\cdot (n\in \Nat\Imp \cpl{b}\cap \fix(\Fwb)\cap v(n)\subseteq \grs(S)\cap S(\m{v'}(n))) &  9 and 1    \\
11 & \forall n\cdot (n\in \Nat\Imp \fix(\Fwb)\cap v(n)\subseteq b\cup \grs(S)\cap S(\m{v'}(n))) &  10\\
12 & \forall n\cdot (n\in \Nat\Imp \fix(\Fwb)\cap v(n)\subseteq \Fm(b)(\m{v'}(n))) &  11 and (\ref{defFm})\\
13 & \fix(\Fwb)\subseteq \fix(\Fm(b)) &  12, 6 and th. \ref{varth}\\
14 & a\subseteq \fix(\Fm(b)) &  13 and 8\\
15 & a\mapsto b\in \Tm &  14 and def. $\Tm$\\
16 & a\mapsto b\in \Lm &  15, equality (\ref{equalMP})
\end{PreuveL}
\eop

\end{document}